\documentclass[a4paper,fleqn,usenatbib]{mnras}
\usepackage{newtxtext,newtxmath}
\usepackage[T1]{fontenc}
\usepackage{ae,aecompl}
\usepackage{amsmath}
\usepackage{amsfonts}
\usepackage{graphicx}
\usepackage{color,xcolor}
\def\mnras{MNRAS}
\def\aap{A$\&$A}
\def\nat{Nature}
\def\apj{ApJ}
\def\apjl{ApJ Letters}
\def\apjs{ApJS}
\def\pasp{PASP}
\def\aj{AJ}

\newcommand{\me}{\, {\rm M}_{\oplus}}
\newcommand{\msun}{\, {\rm M}_{\odot}}
\newcommand{\msunyr}{\, {\rm M}_{\odot}\,{\rm yr^{-1}}}
\newcommand{\rsun}{\, {\rm R}_{\odot}}
\newcommand{\au}{\, {\rm au}}

\title[d203-504 theoretical insights]{Using simultaneous mass accretion and external photoevaporation rates for d203-504 to constrain disc evolution processes}
\author[G. A. L. Coleman et al]{Gavin A. L. Coleman$^{1}$\thanks{Email: gavin.coleman@qmul.ac.uk}, Thomas J. Haworth$^{1}$, Ilane Schroetter$^{2}$ and Olivier Bern\'{e}$^{2}$\\
1. Astronomy Unit, Department of Physics and Astronomy, Queen Mary University of London, Mile End Road, London, E1 4NS, UK\\
2. Institut de Recherche en Astrophysique et Plan\'{e}tologie, Universit\'{e} de Toulouse, CNRS, CNES, 9 Av. du colonel Roche, Toulouse Cedex 4, 31028, France}
\date{Accepted 2025 November 11. Received 2025 October 14; in original form 2025 May 30}
\pubyear{2025}
\begin{document}
\label{firstpage}
\pagerange{\pageref{firstpage}--\pageref{lastpage}}
\maketitle
\begin{abstract}
We cannot understand planet formation without understanding disc evolutionary processes. However, there is currently ambiguity about how protoplanetary discs transport angular momentum (e.g. via viscosity or winds) and the relative contributions and interplay of different dispersal mechanisms. A key difficulty is that for any given system only a handful of disc parameters are usually available to constrain theoretical models. Recent observations of the d203-504 disc in Orion, have yielded values of the stellar accretion rate, external photoevaporative mass loss rate, stellar mass and the disc size and mass. In particular, having the combination of accretion rate and external photoevaporative rate is new. Using this unique combination of observables, we run a suite of disc evolution simulations to constrain which scenarios can match the observed values. We explore both viscous and MHD wind-driven discs, finding that they best match observations when the angular momentum transport $\alpha$ parameter is $3\times10^{-4}\leq\alpha_{\nu}\leq2\times10^{-3}$ for viscous discs, and $2\times10^{-3}\leq\alpha_{\rm DW}\leq10^{-2}$ for MHD wind-driven discs, consistent with other estimates in the literature. As well constraining the disc properties and evolution, the d203-504 disc allows us to define a new irradiation age, since in order to match observations, it was required that the disc had only just appeared in the extreme UV environment it is currently exposed to (a known issue for proplyds referred to as the proplyd lifetime problem). This indicates that it is either very young, i.e. <0.1 Myr, or it has been shielded until recently, which would have protected the planet forming reservoir and helped facilitate planet growth despite it now residing in a harsh UV environment. 

\end{abstract}

\section{Introduction}
\label{sec:intro}

As planets are thought to form in protoplanetary discs surrounding young pre-main sequence stars \citep[e.g.][]{Lissauer1993, JohansenLambrechts2017, Andrews18, Keppler18, Pinte18, Teague19}, it is important to understand the evolution of such discs and the environments in which they inhabit \citep{Lada03, Fatuzzo08, Winter20}. Often, these environments are populated by massive O and B type stars, which give off tremendous amounts of high energy UV radiation that can disperse the star forming cloud \citep[e.g.][]{Mellema06, Dale13, Walch13, Dale15, Bending20, Ali21, Grudic21, Dobbs22}. The UV radiation also strips material from protoplanetary discs, commonly referred to as external photoevaporation \citep[e.g. see][for a  review]{Winter22b}.

From observations, there is now ample evidence showing that external photoevaporation has a significant effect on the evolution of protoplanetary discs \citep[e.g.][for a review]{PastPresentFuture2025}. The most distinct evidence can be found in star forming regions in Orion \citep[$\sim$390 pc distant ][]{MaizApellaniz22}, where photoevaporating discs exhibit cometary-like morphology due to the winds being driven from them, and we refer to these objects as `proplyds' \citep{LaquesVidal1979, Odell93, Odell94, Odell98, Henney99, Bally2000, Odell01, Ricci08, Kim16, Haworth21, Winter22b}. Combining the incident ionising flux and the radius of the ionisation front, one can estimate the mass loss rate from proplyds. These are routinely inferred to be $\sim10^{-6}$\,M$_\odot$\,yr$^{-1}$, which implies disc depletion timescales ($M/\dot{M}$) of order 0.1\,Myr \citep[e.g.][]{Henney99}. This is a lower limit on the true remaining disc lifetime, as the mass loss rate is a function of disc radius and the external far ultraviolet (FUV) radiation field strength, which both change with time \citep{Clarke07, Anderson13,ConchaRamirez19, Qiao22, Wilhelm23} but regardless it illustrates that the inferred mass loss rates are at a level that is significant.

External photoevaporation has also been shown to be influential in theoretical studies of protoplanetary discs, describing the expected mass loss rates \citep{Johnstone98, Adams04, Facchini16, Haworth18, Haworth23} and the associated effect on disc evolution \citep[e.g.][]{Clarke07, Haworth18b, Winter18, Sellek20,ColemanHaworth20,Coleman22, Coleman24MHD, Coleman25, Anania25}, and even more recently on planet formation \citep{2021A&A...656A..72B, Winter22, Qiao23, Huang24, Coleman24,Coleman24FFP}. It was also recently shown to occur for both viscous and MHD wind driven discs, where the effects of external photoevaporation obscure observable differences between the two accretion mechanisms \citep{Coleman24MHD}. Additionally, the UV field that discs experience may not be static, as shielding from the natal cloud \citep{Qiao22,Wilhelm2023} or dynamical movement through a cluster \citep{Coleman25Flyby,Coleman25SigOri}, will both introduce a time varying UV field that can further influence the evolution of protoplanetary discs.

In order to determine the differences between which mechanisms are responsible for the evolution of protoplanetary discs, and to what their underlying properties, a range of observable features are required. These include: accretion rate on to the central star \citep[e.g.][]{Manara16,Manara20,Manara21,Ansdell18,Trapman20} that represents the underlying mode of angular momentum transfer in the discs; disc size and mass \citep[e.g.][]{Mann14,Ansdell17,Eisner18, 2020ApJ...894...74B, VanTerwisga23,Coleman25MDUV} that again provide clues into the both the underlying mechanism for angular momentum transfer, as well as the extent of mass loss through photoevaporative processes, and the ages of systems; and finally the photoevaporative mass loss rate \citep{Odell94, Richling00, Adams04, Smith05, Kim16, Haworth19, Haworth21, Winter22b, 2023ApJ...954..127B, 2025ApJ...983...81B, PastPresentFuture2025}. Whilst currently there are measurements and theoretical estimates of all of these observables, there are few, if any, examples of singular discs with simultaneous precise measurements for all of them. Should such measurements of a single disc be obtained, then that disc could provide valuable insights into the evolution of protoplanetary discs.

In this paper we study the d203-504 disc, which is the first system that we know of with empirical accretion and external photoevaporation rates as well as estimates for the disc mass and the disc size \citep{Schroetter25}. Our goal is to determine and demonstrate the diagnostic power of having both mass loss rates, and the other disc properties, to constrain the evolution of the disc. Additionally, we also aim to establish the approximate age of the disc, and whether it has always been in such a strong UV environment or if it has only just moved into that environment from a more shielded region.

This paper is laid out as follows. We describe the current observations of d203-504 in Sect. \ref{sec:observables}, before outlining our disc model in Sect. \ref{sec:disc_model}. In Sect. \ref{sec:results} we present the results of our simulations outlining where d203-504 sits within our model parameters. We then demonstrate the ability of using the observables of d203-504 to be to place constraints on fundamental disc properties in Sect. \ref{sec:constraints}. Finally we draw our conclusions in Sect. \ref{sec:conc}.

\section{Observables of d203-504} 
\label{sec:observables}

The protoplanetary disc d203-504 is a known externally photoevaporating young stellar object situated near the boundary of the Orion bar, and is associated with the outflow HH 519. \textit{HST} observations of d203-504 have revealed it to have a classic teardrop shaped ``proplyd'' morphology pointing towards  the O9.5V star $\theta^2$ Ori A, indicating that the EUV radiation irradiating the disc originates from there, rather than the O6V star $\theta^1$ Ori C that is associated with most of the proplyds in the region \citep{Bally2000,Odell17}. Observations with VLT/MUSE demonstrated that the disc is almost face on, with a jet oriented almost along the line of sight, and also enabled mass loss rate estimates  \citep{Haworth23VLT, Aru24}. \citep{Haworth23VLT} also concluded that d203-504 was in close proximity to the Orion bar compared to the other nearby evaporating disc d203-506, which shows no sign of an ionisation front and is in silhouette against the Orion bar. 

As we will discuss below, d203-504 has been observed by many different instruments, placing empirical constraints on the disc parameters. The disc is located in the Orion Nebula, a few arcseconds southeast of the ionization front of the Orion Bar \citep{Odell94}. Based on the geometry of the ionisation front, d203-504 is predominantly irradiated by the B star $\theta^2$ Ori A \citep{Bally2000,Odell17}, which is situated at a projected distance of 0.077 pc ($40''$) to the northeast. Using the projected distances to nearby UV sources, \citet{Haworth23VLT} determine an upper limit for the local UV field strength of d203-504 to be $F_{\rm FUV,max}\leq 8\times10^4 \rm G_0$. They also determine the radius of the ionisation front to be equal to $139 \au$, which combined with the projected distance to $\theta^2$ Ori A, and its corresponding ionising luminosity of $3.63\times10^{48}$ photons s$^{-1}$, results in a mass loss rate of $\sim3\times 10^{-7}\msunyr$.

Using MUSE data, \citet{Aru24} determined that the central star for d203-504 is of spectral type K5.5 and had a mass of $0.72\msun$. By using the MUSE data presented in \citet{Haworth23VLT}, \citet{Schroetter25} derived a disc size of $r_{\rm d}=31\au$. Whilst undertaking ALMA observations of the nearby disc d203-506, \citet{Berne24} also obtained 344 GHz dust continuum emissions for d203-504. Analysing that data, \citet{Schroetter25} estimated the mass of the dust disc to be between 4--18$\me$, comparable with other observed dust disc masses \citep{Miotello23}. 

Most recently, \citet{Schroetter25} presented the background subtracted combined \textit{JWST} NIRSpec-MIRI spectrum of  d203-504 as part of the JWST early release science program PDRs4All \citep{Berne22}. They revealed spectroscopic signatures of CO, $\rm H_2$O, C$\rm H_3^+$, and PAHs (polycyclic aromatic hydrocarbons), with the water and CO being detected in absorption in the inner disc regions, close to the central star, indicating that the estimated gas-phase C/O ratio is $\sim0.48$, consistent with both the Solar value, and that observed in the Orion Nebula. They also found  that the spectrum is characterized by a strong near-infrared continuum emission that can be well fitted by a 1180 K blackbody, with a luminosity $L_{\rm NIR} = 0.13 L_{\bigodot}$. By assuming that viscous heating is the main contributor to the luminosity in the inner disc region close to the central star, and that the emission there is optically thick, this allowed \citet{Schroetter25} to be able to derive an accretion through disc, corresponding to $\dot{M}_{\rm acc}=8.2\times 10^{-9}\msunyr$, that is consistent with accretion rates observed around other young T-Tauri stars \citep{Rigliaco13}. Additionally, by looking at the line intensity of $Pa_{\alpha}$in \citet{Schroetter25}, we can compute a $Pa_{\alpha}$ luminosity, $L_{Pa_{\alpha}}=4\times10^{-4}L_{\odot}$, which corresponds to $L_{\rm acc}=0.032 L_{\odot}$ following \citet{Rogers24}. This equates to an accretion rate $\dot{M}_{\rm acc}=2\times 10^{-8}\msunyr$, similar to that determined by \citet{Schroetter25} from the near-infrared continuum emission. The middle section of Table \ref{tab:parameters} shows the observational constraints for d203-504.

With observations providing both the accretion rate and the photoevaporation rate, this means that d203-504 is the first disc where the main mass loss mechanisms are simultaneously known. Coupling the estimated mass loss rates, with the disc size and mass allows us to test how well disc evolution models can match the observations, giving insights into how protoplanetary discs evolve. We will address this in the following sections.

\section{Disc Model for d203-504}
\label{sec:disc_model}

Protoplanetary discs lose mass by accretion onto the central star and through photoevaporative winds launched from the disc surface layers. To account for these, it is necessary to use a 1D protoplanetary disc model that accounts for the exchange of angular momentum through the disc, whilst also including the loss of gas in the outer disc regions. We utilise the model presented in \citet{Coleman24MHD} that includes both a 1D viscous $\alpha$ disc model \citep{Shak}, as well as contributions from MHD disc winds \citep{Tabone22}. Both viscosity and MHD winds allow for accretion through the disc and on to the central star. In addition to this accretion, we also include the mass loss of the disc from both internal \citep{Picogna21} and external photoevaporation \citep{Haworth23}. We will briefly describe the model below.

To evolve the gas surface density $\Sigma$ of the disc, we solve the diffusion equation, taking into account viscous accretion, MHD winds, and the mass loss from photoevaporation
\begin{equation}
\label{eq:diffusion}
\begin{split}
    \dot{\Sigma}(r)=&\dfrac{1}{r}\dfrac{d}{dr}\left[3r^{1/2}\dfrac{d}{dr}\left(\nu\Sigma r^{1/2}\right)\right]+\dfrac{3}{2r}\dfrac{d}{dr}\left[\dfrac{\alpha_{\rm DW}\Sigma c_{\rm s}^2}{\Omega}\right]\\
    &-\dfrac{3\alpha_{\rm DW}\Sigma c_{\rm s}^2}{4(\lambda-1)r^2\Omega}-\dot{\Sigma}_{\rm PE}(r)
\end{split}
\end{equation}
where $\nu=\alpha_{\rm v} H^2\Omega$ is the disc viscosity with viscous parameter $\alpha_{\rm v}$ \citep{Shak}, $H$ is the disc scale height and $\Omega$ the Keplerian frequency.
The second and third components on the right hand side of eq. \ref{eq:diffusion} represent the change in surface density due to the angular momentum extracted by MHD wind, and the the mass lost in the MHD wind itself \citep[e.g.][]{Pascucci25, Schwarz25, Arulanantham24}, where $\lambda$ is the magnetic lever arm parameter that quantifies the efficiency of the wind to carry out angular momentum \citep{Blandford82}. We assume that $\lambda=3$ for the simulations presented in this work, consistent with observationally derived estimates that find $\lambda\simeq$2--6 \citep{Tabone17,DeValon20,Booth21,Nazari24}. Note that whilst we use $\lambda=3$ for the simulations presented in this work, the effects of changing the lever arm between 2--6 are presented in Appendix \ref{appendix:lever}.
The fourth term represents mass extracted by photoevaporative winds, which we discuss below.
Note there are two components for $\alpha$ in eq. \ref{eq:diffusion}, those being the measure of turbulence for viscosity $\alpha_{\rm v}$, and the measure of angular momentum extracted in the wind $\alpha_{\rm DW}$. In this work, we will explore discs that are either viscous, or MHD wind driven, with one parameter taking the full value of $\alpha$, and the other being set to 0.

To account for photoevaporative winds, we follow \citet{Coleman22} and assume that the photoevaporative mass loss rate at any radius of the disc is the maximum of the internally and externally driven rates 
\begin{equation}
    \dot{\Sigma}_{\rm PE}(r) ={\rm max}\left(\dot{\Sigma}_{\rm I,X}(r),\dot{\Sigma}_{\rm E,FUV}(r)\right)
\end{equation}
where the subscripts I and E refer to contributions from internal and external photoevaporation.
This allows us to account for both the high energy X-ray photons emanating from the central star that launch internally driven photoevaporative winds, as well as winds launched from the outer disc by far ultraviolet (FUV) radiation emanating from nearby massive stars (e.g. O-type stars).

For internally driven winds, we follow \citet{Picogna21} who further build on the work of \cite{Picogna19} and \cite{Ercolano21} and find that the mass loss profile from internal X-ray irradiation is approximated by
\begin{equation}
\label{eq:sig_dot_xray}
\begin{split}
\dot{\Sigma}_{\rm I,X}(r)=&\ln{(10)}\left(\dfrac{6a\ln(r)^5}{r\ln(10)^6}+\dfrac{5b\ln(r)^4}{r\ln(10)^5}+\dfrac{4c\ln(r)^3}{r\ln(10)^4}\right.\\
&\left.+\dfrac{3d\ln(r)^2}{r\ln(10)^3}+\dfrac{2e\ln(r)}{r\ln(10)^2}+\dfrac{f}{r\ln(10)}\right)\\
&\times\dfrac{\dot{M}_{\rm X}(r)}{2\pi r} \dfrac{\msun}{\au^2 {\rm yr}}
\end{split}
\end{equation}
where
\begin{equation}
\label{eq:m_dot_r_xray}
    \dfrac{\dot{M}_{\rm X}(r)}{\dot{M}_{\rm X}(L_{X})} = 10^{a\log r^6+b\log r^5+c\log r^4+d\log r^3+e\log r^2+f\log r+g}.
\end{equation}
For a 0.7$\msun$ star similar to d203-504, \citet{Picogna21} found that $a=0.3034$, $b=-1.5323$, $c=1.5766$, $d=4.0211$, $e=-11.1311$, $f=10.655$, and $g=-4.5769$.

Following \cite{Ercolano21} the integrated mass-loss rate, dependent on the stellar X-ray luminosity, is given as
\begin{equation}
    \log_{10}\left[\dot{M}_{X}(L_X)\right] = A_{\rm L}\exp\left[\dfrac{(\ln(\log_{10}(L_X))-B_{\rm L})^2}{C_{\rm L}}\right]+D_{\rm L},
\end{equation}
in $\msunyr$, with $A_{\rm L} = -1.947\times10^{17}$, $B_{\rm L} = -1.572\times10^{-4}$, $C_{\rm L} = -0.2866$, and $D_{\rm L} = -6.694$. By including the additional cooling effects due to the excitation of O from neutral H, \citet{Sellek24} found that the mass loss rates found in \citet{Picogna21} are overestimated by factor $\sim9$, and as such we apply this correction to the equation above.

In addition to internal winds driven by irradiation from the host star, winds can also be driven from the outer regions of discs by irradiation from external sources. Massive stars dominate the production of UV photons in stellar clusters and hence dominate the external photoevaporation of discs. External photoevaporation has been shown to play an important role in setting the evolutionary pathway of protoplanetary discs \citep[e.g.][]{Clarke07, Haworth18b, Winter18, Sellek20,Coleman22, Coleman24MHD, Coleman25, Anania25}, their masses \citep{Mann14,Ansdell17,VanTerwisga23}, radii \citep[e.g.][]{Eisner18,2020ApJ...894...74B} and lifetimes \citep{2015A&A...578A...4S, Guarcello16,ConchaRamirez19,Sellek20, 2022ApJ...939L..10P} even in weak UV environments \citep{Haworth17}.

\begin{table}
    \centering
    \begin{tabular}{c|c}
    \hline
   Fixed Simulation Parameters & Value \\
    \hline
        $r_{\rm in} (\au)$ & 0.036 \\
        $r_{\rm out} (\au)$ & 500 \\
        $M_{*} (\msun)$ & 0.72 \\
        $T_* (K)$ & 4100 \\
        $R_* (\rsun)$ & 1.85 \\
        $L_X (\log_{10}(\rm erg~s^{-1}))$ & 29.5 \\
        $F_{\rm FUV,0} (\rm G_0)$ & 10 \\
        $F_{\rm FUV,max} (\rm G_0)$ & $8\times10^4$ \\
        \hline
        Observational Constraints\\
        \hline
        $r_{\rm d}$ & 31$\au$ \\
        $M_{\rm dust}$ & 4--18$\me$\\
        $\dot{M}_{\rm E,FUV} (\msunyr)$ & $3\times10^{-7}$\\
        $\dot{M}_{\rm acc} (\msunyr)$ & $8.2\times10^{-9}\msunyr$\\
        \hline
        Variable Simulation Parameters & Value \\
        \hline
        $M_{\rm disc} (M_{\rm d,max})$ & 0.02, 0.06, 0.1, ..., 0.96, 1\\
        $r_{\rm C} (\au)$ & 20, 30, 40, 50\\
        $t_{\rm sh}$ (Myr) & 0, 0.1, 0.2, ..., 1.4, 1.5\\
        $\alpha_{\nu}$ & $10^{-4}$, $10^{-3.9}$, $10^{-3.8}$, ..., $10^{-2.6}$, $10^{-2.5}$\\
        $\alpha_{\rm DW}$ & $10^{-3}$, $10^{-2.9}$, $10^{-2.8}$, ..., $10^{-1.6}$, $10^{-1.5}$\\
    \hline
    \end{tabular}
    \caption{Simulation Parameters for the models discussed in this paper.}
    \label{tab:parameters}
\end{table}

In our simulations, the mass loss rate due to external photoevaporation is calculated by interpolating over the recently updated \textsc{fried} grid \citep{Haworth23}.
This new grid expands on the original version of \textsc{fried} \citep{Haworth18} in terms of the breadth of parameter space in UV field, stellar mass, disc mass and disc radius. The \textsc{fried} grid provides mass loss rates for discs irradiated by FUV radiation as a function of the star/disc/FUV parameters. In our simulations, we determine the mass loss rate at each time step by linearly interpolating \textsc{fried} in three dimensions: disc size $r_{d}$, disc outer edge surface density $\Sigma_{\textrm{out}}$ and FUV field strength $F_{\rm{FUV}}$.

We evaluate the \textsc{fried} mass loss rate at each radius from the outer edge of the disc down to the radius that contains 80$\%$ of the disc mass. We choose this value as 2D hydrodynamical models show that the vast majority of the mass loss from external photoevaporation, comes from the outer 20\% of the disc \citep{Haworth19}. The change in gas surface density is then calculated as
\begin{equation}
    \dot{\Sigma}_{\textrm{E, FUV}}(r) = G_{\rm sm} \frac{\dot{M}_\textrm{{ext}}(r_{\textrm{\textrm{max}}})}{\pi(r^2_\textrm{{d}} - {r_{\textrm{\textrm{max}}}}^2)+A_{\rm sm}}, 
\end{equation}
where $A_{\rm sm}$ is a smoothing area equal to 
\begin{equation}
A_{\rm sm} = \dfrac{\pi(r_{\rm max}^{22}-(0.1 r_{\rm max})^{22})}{11r_{\rm max}^{20}}
\end{equation}
and $G_{\rm sm}$ is a smoothing function
\begin{equation}
    G_{\rm sm} = \dfrac{r^{20}}{r_{\rm max}^{20}}.
\end{equation}
The outer disc radius $r_{\rm d}$ is calculated as the outermost location where the surface density $\Sigma(r)>10^{-4} \rm gcm^{-2}$, whilst $r_{\rm max}$ is the radial location where the mass-loss rate is at a maximum. We include the smoothing function so as not to introduce discontinuities into the numerical setup. The choice of the smoothing function does not affect the overall evolution of the discs, compared to other approaches for implementing the mass loss rates from \textsc{fried} \citep[e.g.][]{Sellek20}.

\begin{figure*}
\centering
\includegraphics[scale=0.5]{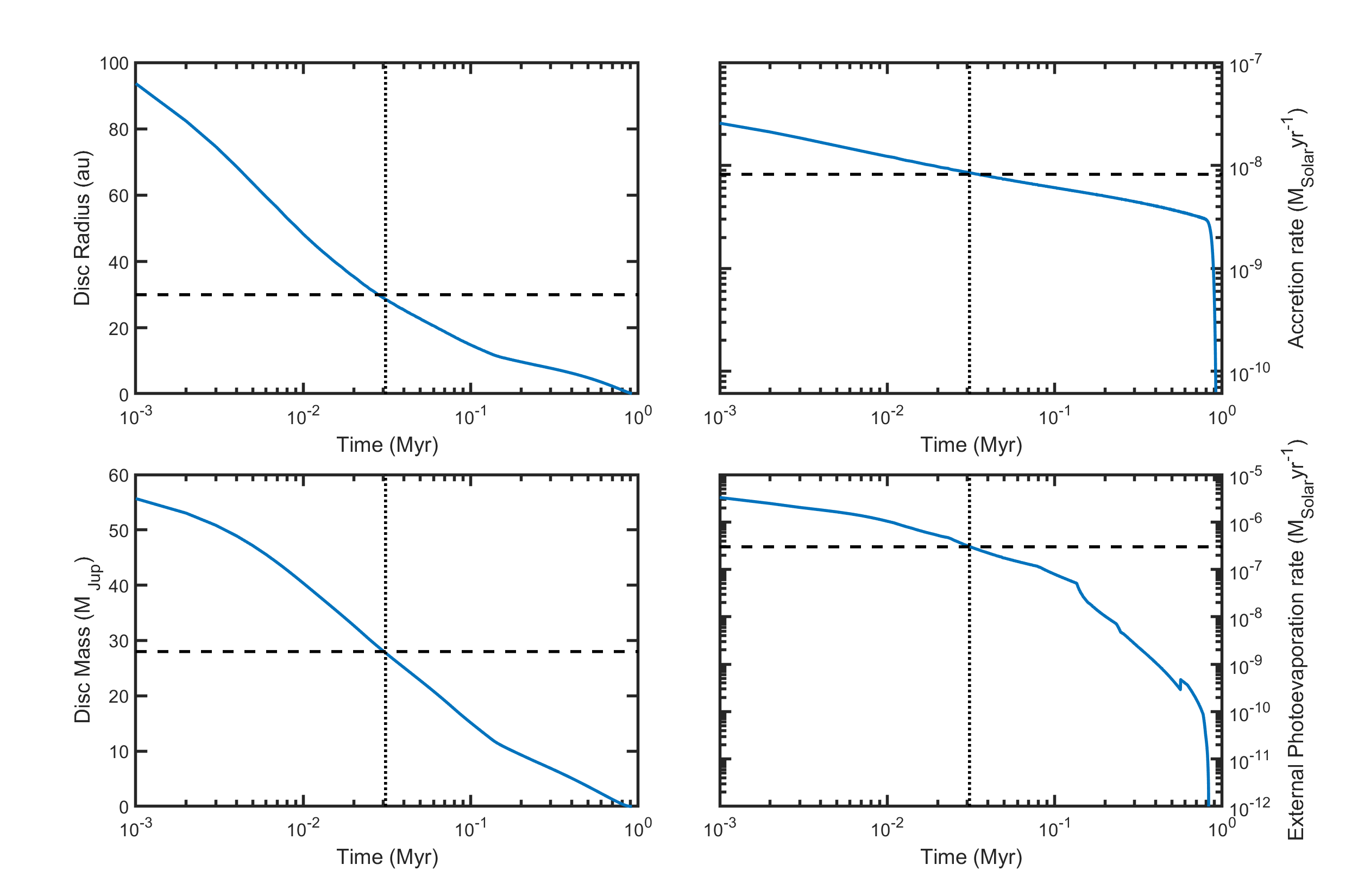}
\caption{Temporal evolution of the disc radius (top-left), disc mass (bottom-left), accretion rate (top-right) and the external photoevaporation rate (bottom-right) for the simulation that best fits the observed values. The disc had a shielding time $t_{\rm sh}=0$ Myr, the initial scale radius $r_{\rm C}=50\au$, the initial disc mass $M_{\rm d}=0.057\msun$, and $\alpha=3\times10^{-3}$. The dashed horizontal line denotes the observed values for d203-504, whilst the dotted vertical line shows the time at which the simulation best fits the observations.}
\label{fig:example_best}
\end{figure*}

The {\sc fried} grid contains multiple subgrids that vary the PAH-to-dust ratio ($\rm f_{PAH}$) and specify whether or not grain growth has occurred.
The effects of using different combinations of these parameters will be explored in future work, but we do not expect such changes to affect the differences between viscous and MHD wind driven discs.
The combination we use here is $\rm f_{PAH}=1$ (an interstellar medium, ISM,-like PAH-to-dust ratio) and assume that grain growth has occurred in the outer disc, depleting it and the wind of small grains which reduces the extinction in the wind and increases the mass loss rate compared to when dust is still ISM-like. This results in a depleted PAH-to-gas ratio of 1/100, which was recommended as fiducial by \citet{Haworth23, Coleman25} and is motivated also by \cite{Vicente13} find signs of PAH depletion in the proplyd HST 10. However, a recent estimate of the PAH abundance in d203-504 suggests a PAH-to-gas ratio closer to the standard ISM value, with a factor $\sim16$ depletion \citep{Schroetter25}. We discuss this in section \ref{sec:PAHtogas}.

Recent work has shown that as stars form in stellar clusters, they may be initially located in shielded environments, where the surrounding gas cloud reduces the number of penetrating UV photons that impact the discs and launch photoevaporative winds \citep{Qiao22,Qiao23, Wilhelm23}. We investigate the influence of shielding here on the evolution of d203-504 by utilising the prescriptions found in \citet{Qiao23}. These prescriptions use a parameterised time varying FUV flux profile that transitions from a low FUV field $F_{\textrm{FUV, 0}}$ to a high value $F_{\textrm{FUV, max}}$ after a shielding period $t_{\textrm{sh}}$
\begin{multline}
\label{eqn:FUV_track}
    F_{\rm FUV}(t) = F_{\rm FUV,0} + \frac{1}{2}(F_{\rm FUV,max}  - F_{\rm FUV,0}) \\ 
    \times \left(\tanh \left(\frac{t - t_\textrm{{sh}}}{t_{\rm trans}}\right) + 1\right),
\end{multline}
where $t_{\rm trans}$ is a parameter that controls the time-scale over which the star/disc transitions from a shielded to an unshielded environment. This rapid transition represents the change in irradiation when stars/discs cease being embedded in the cloud \citep{Qiao22}. We set $F_{\rm FUV,0} = 10 \rm G_0$, and $F_{\rm FUV, max}$ to $8\times10^4 \rm G_0$, that observed for d203-504 \citep{Haworth23VLT}. To explore the effects of shielding we vary the shielding time and the transition time between a shielded and unshielded environment. Note, that with the FUV field strength being so high, external photoevaporation dominates over internal photoevaporation in determining the evolution of the discs explored here consistent with expectations from previous works \citep[e.g.][]{Coleman22}. With external photoevaporation dominating the disc evolution, our results would be effectively unchanged should internal photoevaporation not be included.

\begin{figure}
\centering
\includegraphics[scale=0.55]{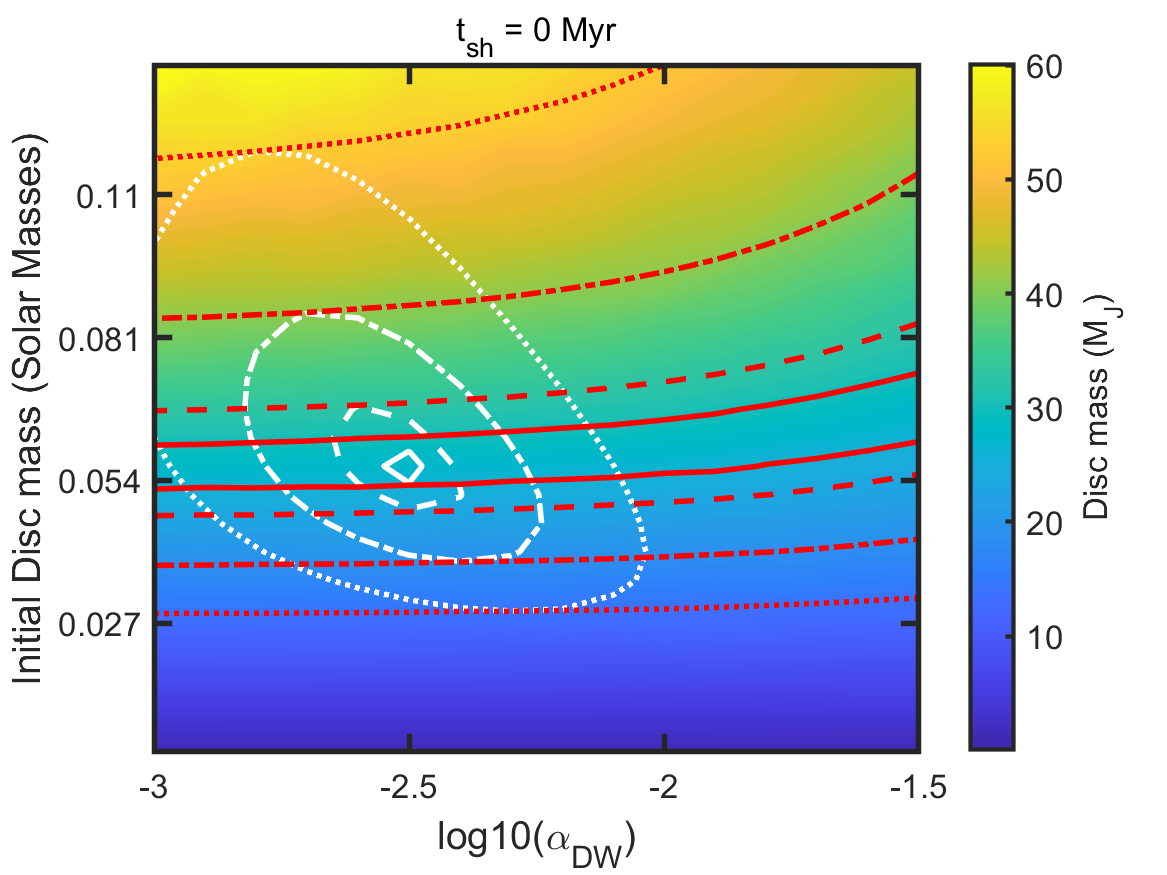}
\caption{Contour plot showing the disc mass for at the time of best fit to the observed data for d203-504. The white contours highlight regions where the differences in log space between the simulations and observations are less than 0.041 (solid), 0.079 (dashed), 0.176 (dot-dashed) and 0.3 (dotted). These values correspond to differences of factors: 1.1, 1.2, 1.5 and 2 respectively. The red lines shows the same as the white lines, except the observed accretion rates are not compared with the simulations, highlighting the importance of knowing the accretion rate for determining compatible values of $\alpha$.}
\label{fig:contour_wind_50_0}
\end{figure}

\subsection{Simulation Parameters}
To determine the best fit parameters for the observed data, we vary a number of parameters, specifically those that strongly affect the mass accretion and photoevaporative mass loss rates. These include the initial disc mass, its compactness, and the strength of the viscous $\alpha$ turbulent parameters or the MHD wind strength. We initialise our discs following \citet{Lynden-BellPringle1974}
\begin{equation}
    \Sigma = \Sigma_0\left(\frac{r}{1\au}\right)^{-1}\exp{\left(-\frac{r}{r_{\rm C}}\right)}
\end{equation}
where $\Sigma_0$ is the normalisation constant set by the total disc mass, and $r_{\rm C}$ is the scale radius, effectively representing the initial compactness of the disc. To explore this compactness, we vary $r_{\rm C}$ between 20--50$\au$.
\citet{Haworth20} found that the maximum mass a disc could be before it went gravitationally unstable was equal to
\begin{equation}
    \label{eq:max_disc_mass}
    \dfrac{M_{\rm d, max}}{M_*} < 0.17 \left(\dfrac{r_{\rm ini}}{100\au}\right)^{1/2}\left(\dfrac{M_*}{\msun}.\right)^{-1/2}
\end{equation}
Given that we initialize the disc with the \citet{Lynden-BellPringle1974} self-similar solution, and not a constant slope, we take the initial disc radius to be equal to 1.8 $\times$ the scale radius $r_{\rm C}$, in order to obtain appropriate protoplanetary disc masses that are gravitationally stable around a 0.7$\msun$ star. To determine the strength of the initial disc mass, we explore values between 0.02--1 $\times M_{\rm d, max}$. These correspond to initial disc masses being between $1.7\times10^{-3}$--$0.135\msun$. In addition to the initial sizes and masses of the protoplanetary discs, we also examine the differences between viscous and MHD wind driven discs where we vary the strengths of $\alpha_{\rm visc}$ and $\alpha_{\rm dw}$.
For viscous discs, we evolve the discs with $\alpha$ values between $10^{-4}$--$10^{-2.5}$, comparable to values expected from observations \citep[see e.g. ][ for a review]{Lesur23}, and theoretical expectations \citep{Coleman22,Coleman24}.
Since the mass accretion rates for MHD wind driven discs are typically a factor few lower for discs of similar mass and sizes \citep{Coleman24MHD}, we explore values of $\alpha_{\rm dw} = 10^{-3}$--$10^{-1.5}$, an order of magnitude larger than what we explore for $\alpha_{\rm visc}$. These values for $\alpha$ should allow us to explore the parameter space where both viscous and MHD wind driven discs can best match the observations of d203-504. As well as the underlying disc properties, we also vary the shielding times of the discs, exploring between $t_{\rm sh}=0$--$1.5$ Myr. The top section of Table \ref{tab:parameters} shows the parameters that are fixed in this work, whilst the bottom section of Table \ref{tab:parameters} shows the values for the parameters that we vary across the simulations.

\section{Fitting parameters to d203-504}
\label{sec:results}

The main objectives of this work is to determine the disc parameters that are able to match the observational constraints on the properties of d203-504 \citep{Schroetter25}. These include the accretion rate on to the star, $\dot{M}_{\rm acc} = 8.2\times10^{-9}\msunyr$, the external photoevaporation rate, $\dot{M}_{\rm E,FUV}=3\times10^{-7}\msunyr$, the disc radius $r_{\rm d}=31\au$, and a gas disc mass equal to 25 $M_{\rm Jup}$. In order to calculate the disc parameters that best match observations, we define a distance metric
\begin{equation}
\label{eq:distance}
    d^2 = \sum^N_{i=1} (\log_{10}(X_i) - \log_{10}(O_i))^2
\end{equation}
where $X_i$ is the value from the simulations, and $O_i$ is the observed value, with different $i$ values representing the four observational constraints mentioned above, and $N=4$ being the total number of constraints. We take the logarithms of all values to ensure they are all compared on equal footing in regards to their relative differences. We calculate this metric every 1000 years, and take the smallest distance as the best fit value.

\begin{figure*}
\centering
\includegraphics[scale=0.35]{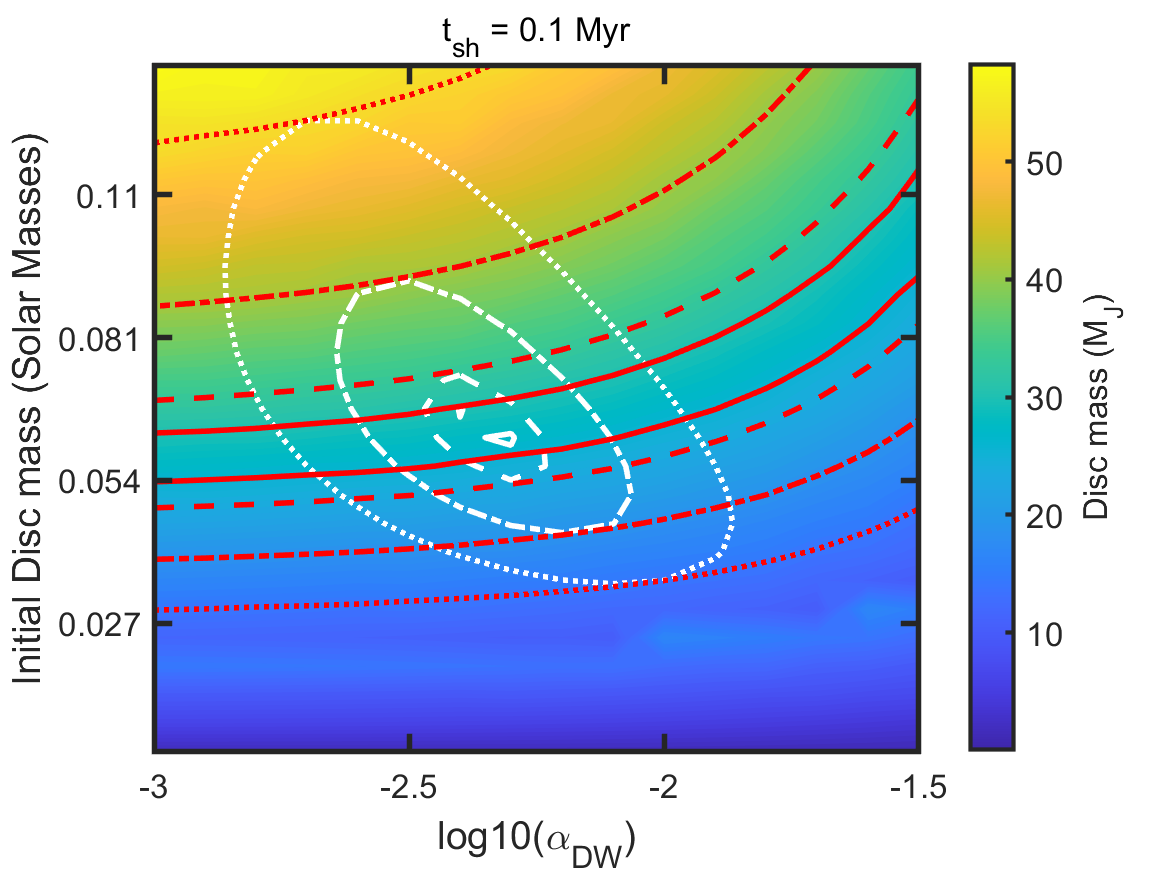}
\includegraphics[scale=0.35]{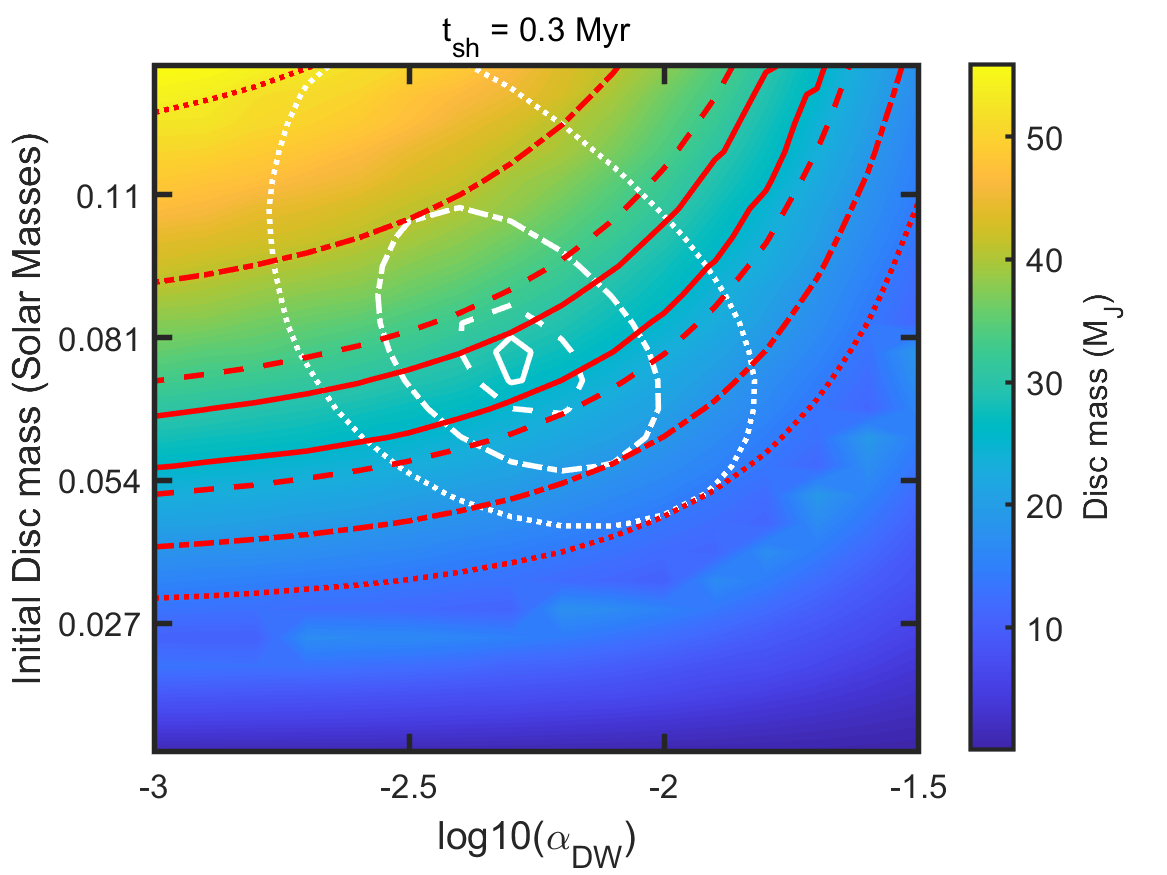}
\includegraphics[scale=0.35]{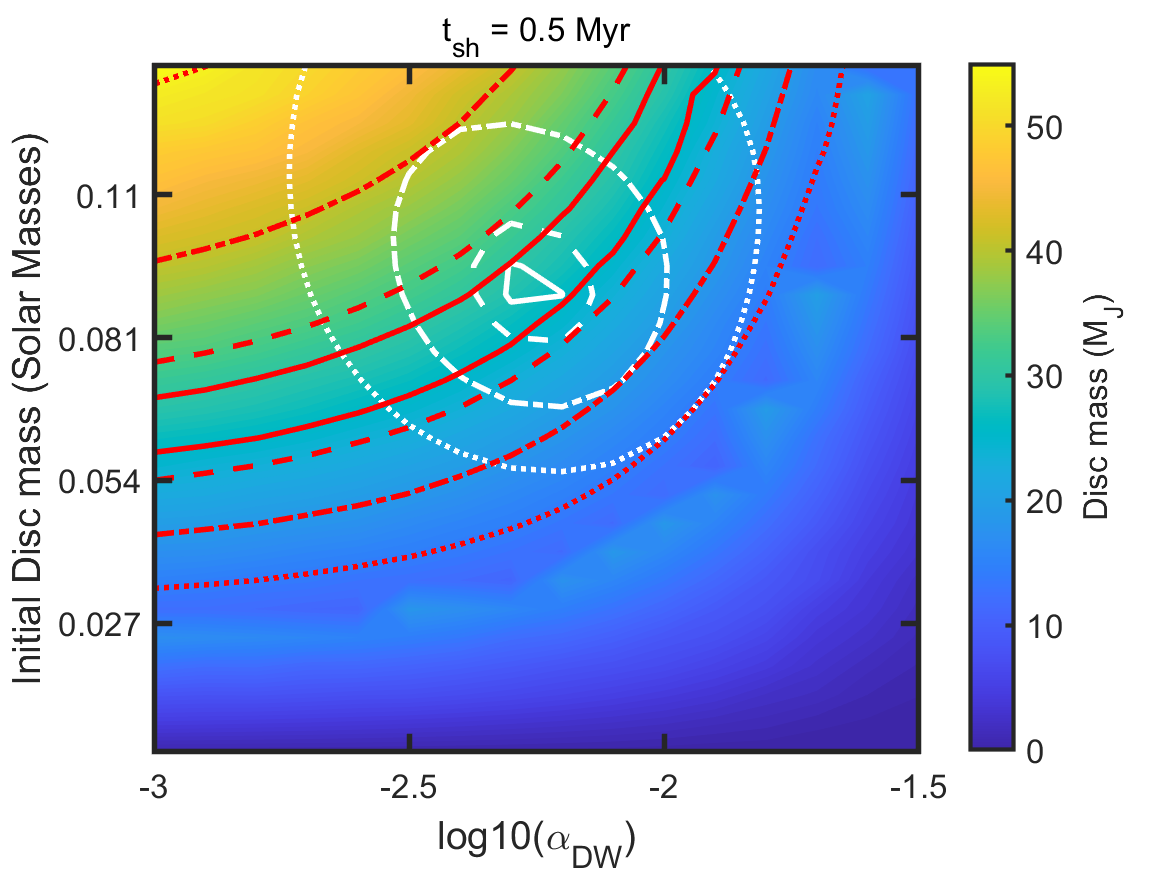}
\caption{Same as Fig. \ref{fig:contour_wind_50_0}, but for different shielding times: 0.1 Myr (left-hand panel), 0.3 Myr (middle panel), and 0.5 Myr (right-hand panel).}
\label{fig:contour_wind_all}
\end{figure*}

\begin{figure*}
\centering
\includegraphics[scale=0.55]{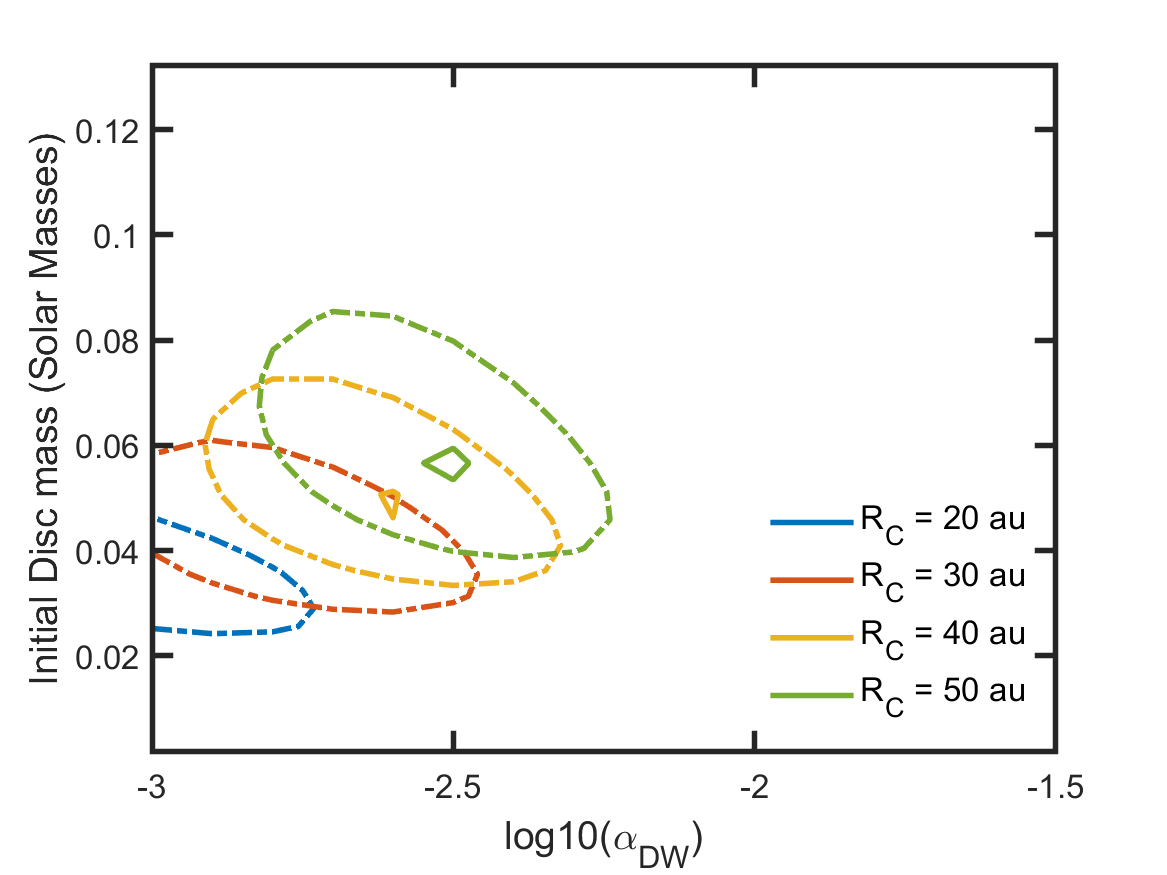}
\includegraphics[scale=0.55]{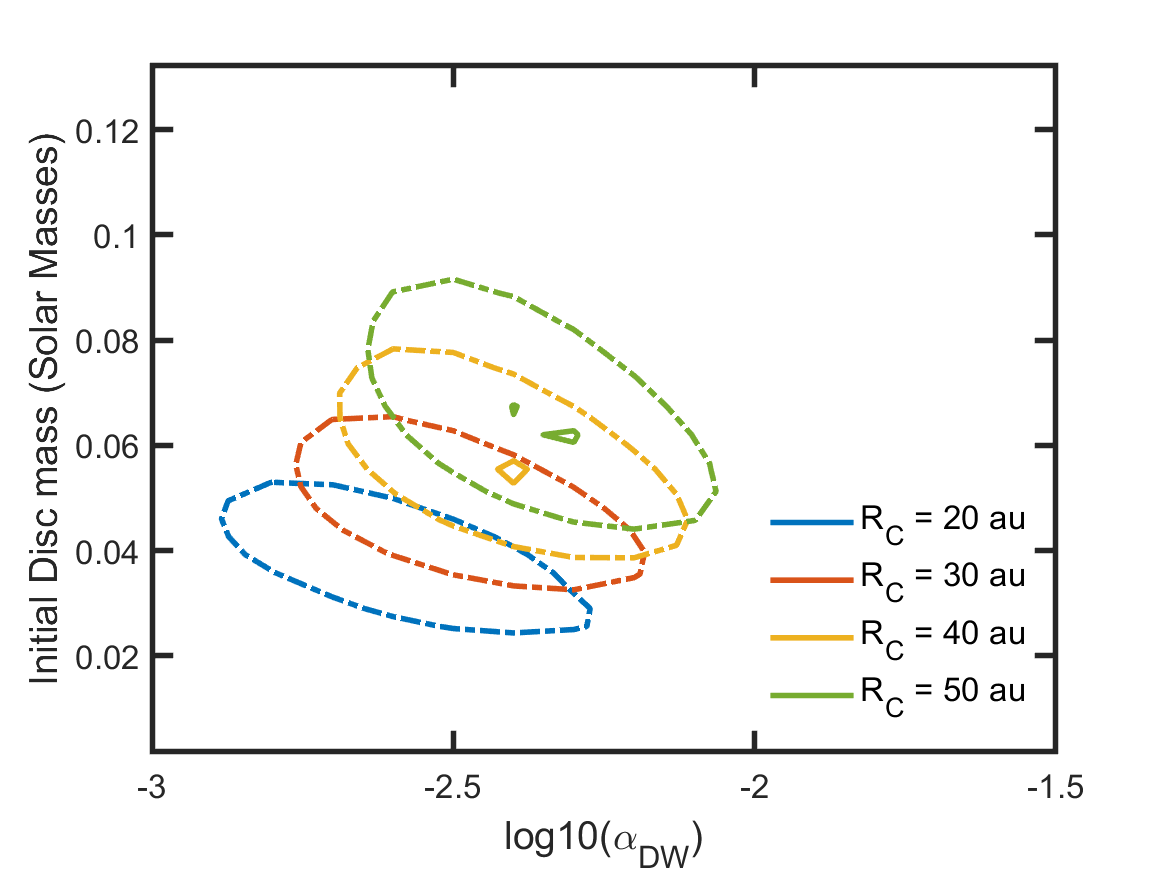}\\
\includegraphics[scale=0.55]{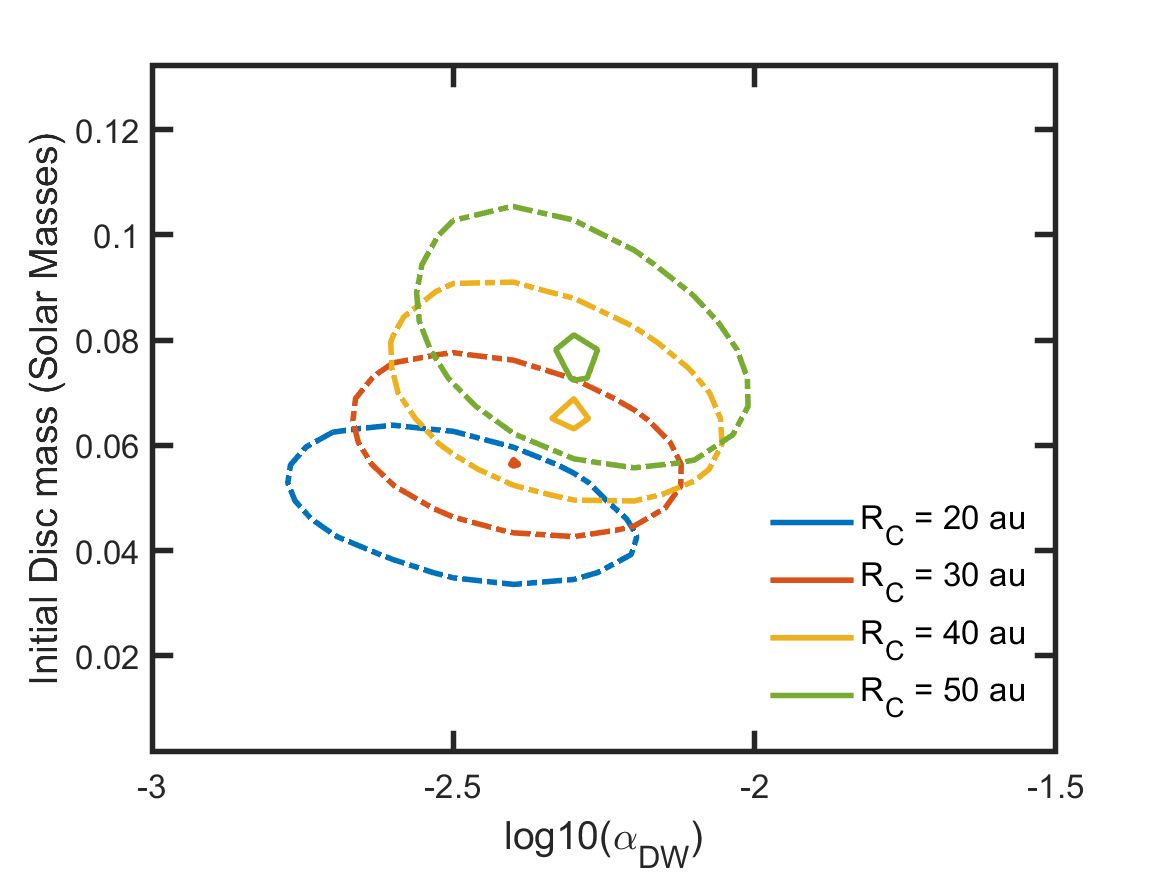}
\includegraphics[scale=0.55]{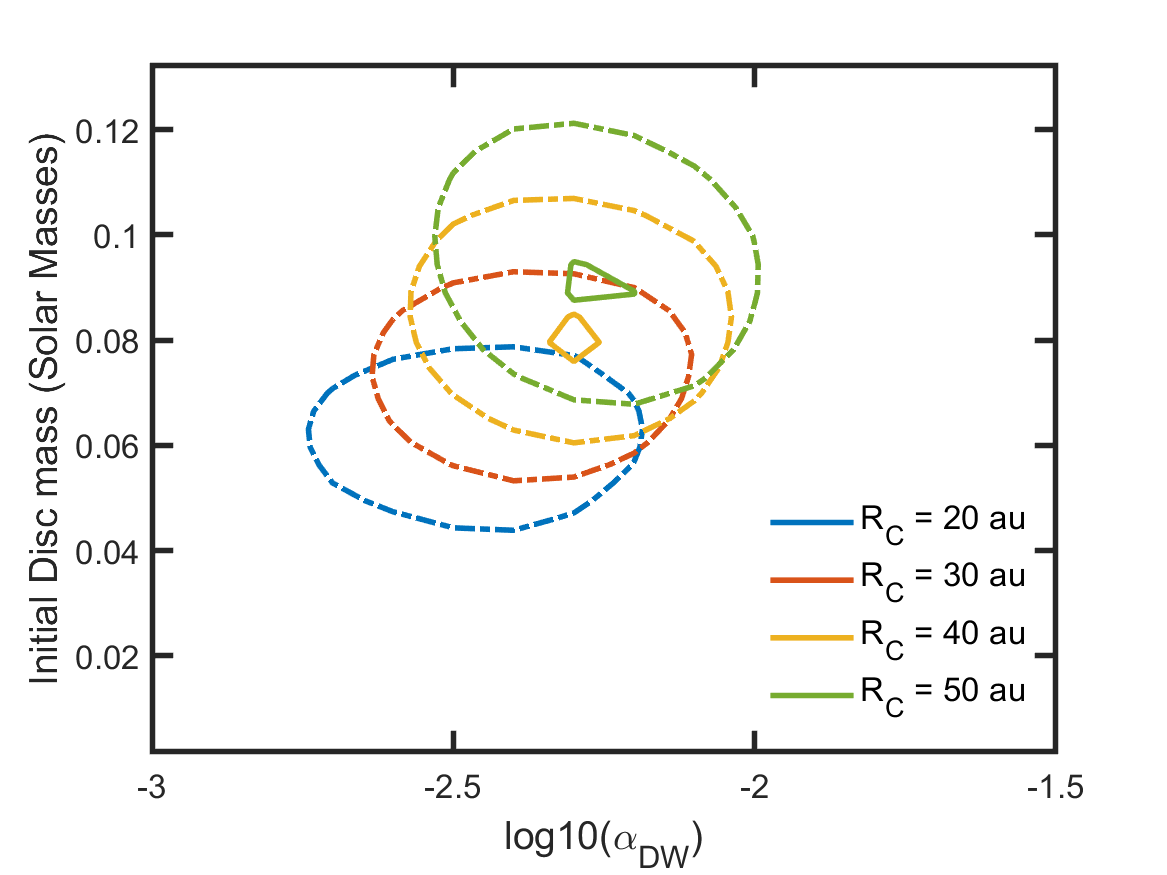}
\caption{Contours showing the locations of simulations that best fit observations for different $\alpha_{\rm DW}$ and disc masses. Different colours show different initial scale radii of 20$\au$ (blue), 30$\au$ (red), 40$\au$ (yellow) and 50$\au$ (green). Solid contours show regions where the simulation matched observations to within a factor 1.1 whilst dashed-dotted contours show for differences of factor 1.5. The different panels differentiate between different shielding times of: 0 Myr (top-left), 0.1 Myr (top-right), 0.3 Myr (bottom-left) and 0.5 Myr (bottom-right).}
\label{fig:alpha_rc_wind}
\end{figure*}

\subsection{MHD Wind discs}
\label{sec:wind}
We first explore whether MHD wind discs are able to match the observed parameters of d203-504. Figure \ref{fig:example_best} shows the evolution of a single disc model that was able to best fit the observed values. The figure shows the temporal evolution of the disc radius, always taken as the radius that encompasses 90\% of the disc mass (top-left), disc mass (bottom-left), mass accretion rate (top-right), and the eternal photoevaporation rate (bottom-right). The dashed horizontal lines denote the observed values, whilst the dotted vertical lines show the time at which the simulation best matches the observed values. The disc had no shielding time, whilst it's initial disc mass was equal to $0.057 \msun$, the initial scale radius was 50$\au$, and the MHD wind $\alpha=3\times10^{-3}$. As there is no shielding, the external radiation quickly reduces the disc in both size and mass, as can be seen on the left-hand panels of Fig. \ref{fig:example_best}. This fast removal of gas resulted in external photoevaporation rates $\ge 10^{-6} \msunyr$ as is shown in the bottom-right panel. After only 0.03 Myr the evolution of the disc had reduced the disc sufficiently in size and mass, with the external photoevaporation rate also dropping by an order of magnitude to being only $\sim3\times10^{-7}\msunyr$, similar to that observed for d203-504. Additionally at this time, 0.03 Myr, the simulated disc radii, mass and accretion rate were also similar to those observed for d203-504. For the simulated disc, it continued to evolve over the next 0.9 Myr, before being fully dispersed. The short lifetime highlights the importance and strength of external UV environments in determining the evolution of protoplanetary discs, similar to that found in disc evolution studies.

Whilst Fig. \ref{fig:example_best} shows that the observed values for d203-504 could be matched from disc evolution models, it does so only for a single set of initial disc parameters. Exploring the initial disc parameters further, Fig. \ref{fig:contour_wind_50_0} shows the disc mass at the time of best fit, i.e. those with the smallest distance from eq. \ref{eq:distance}, for simulations that were initiated with different initial disc masses (vertical-axis) and values of $\alpha_{\rm DW}$ (horizontal-axis). The white lines highlight regions where the differences in log space between the simulations and observations are less than 0.041 (solid), 0.079 (dashed), 0.176 (dot-dashed) and 0.3 (dotted). These values correspond to differences of factors: 1.1, 1.2, 1.5, and 2 respectively. The red lines shows the same as the white lines, except the observed accretion rates are not included when comparing the observations with simulations. All of the discs simulated in Fig. \ref{fig:contour_wind_50_0} had no shielding time, and initial scale radii of 50$\au$. As can be seen by the solid white and red lines, they occupy a small region of the parameter space where the all of the observables could be best matched at some point over the lifetimes of the simulated discs. For other regions of the parameter space, they were unable to match the observables adequately well, with their best fits determined with large errors on some of the parameters, e.g. the disc mass. This can be clearly seen at the top of Fig. \ref{fig:contour_wind_50_0} where the best fit for those discs occurred with disc masses $>50M_{\rm Jup}$.

Looking at the white contours in Fig. \ref{fig:contour_wind_50_0}, the discs that best fit the observed values, shown by the solid white line are found at around an MHD wind $\alpha_{\rm DW}\simeq 10^{-2.5}$, and with initial disc masses of $\sim0.05\msun$. These comparisons included the newly determined observable accretion rates \citep{Schroetter25}, the first determined for proplyds. Without the observed accretion rate, as shown by the red lines, then no constraints are able to be placed on the strength of $\alpha_{\rm DW}$. Indeed the red solid line in Fig. \ref{fig:contour_wind_50_0} is able to match the other observables for a wide range of $\alpha_{\rm DW}$, but for a small range of the initial disc mass. The consistent value of the initial disc mass is due to there being no shielding time, and the best fits occurring at the beginning of the disc lifetimes where they are still extended and able to provide strong external photoevaporative winds.

\subsubsection{Importance of shielding time}

With stars and protoplanetary discs forming in stellar clusters, it may be expected that some of the natal cloud may shield the forming discs as they evolve \citep{Qiao22, Qiao23, Wilhelm23}. Given that the models suggest that the most probable scenario is only recent irradiation, it is possible that d203-504 may have just exited from the Orion Bar, which is coincident with the line of sight, and where d203-504 would have been shielded from the high energy radiation. \citet{Haworth23VLT} do place d203-504 between the near ionised layer and the Orion bar. 

Fig. \ref{fig:contour_wind_all} shows the effects of the shielding time on the contours of best fit. Whilst the best fit values for $\alpha_{\rm DW}$ remain between 3--6$\times10^{-3}$, the disc masses are now slightly increased, since more mass is needed for longer to match the observed accretion rates when the external photoevaporation rates increase as the disc becomes unshielded. However, whilst the white contours (which account for the accretion rate) show little change in $\alpha_{\rm DW}$ as a function of the shielding time, the red contours (everything but the accretion rate) show a different behaviour where the calculated best fit now increases in mass as $\alpha_{\rm DW}$ increases. This is especially the case in discs with larger shielding times, where now some values of $\alpha_{\rm DW}$ require initial disc masses much larger than is gravitationally stable, in order to contain enough mass to match observations when the discs emerge from their shielded environments. These discs would be located above our grid of explored disc masses and those presented in Fig. \ref{fig:contour_wind_all} and are typically at larger $\alpha_{\rm DW}$ values, e.g. $\alpha_{\rm DW}>10^{-2}$ for $t_{\rm sh} = 0.5$Myr.

\begin{figure}
\centering
\includegraphics[scale=0.55]{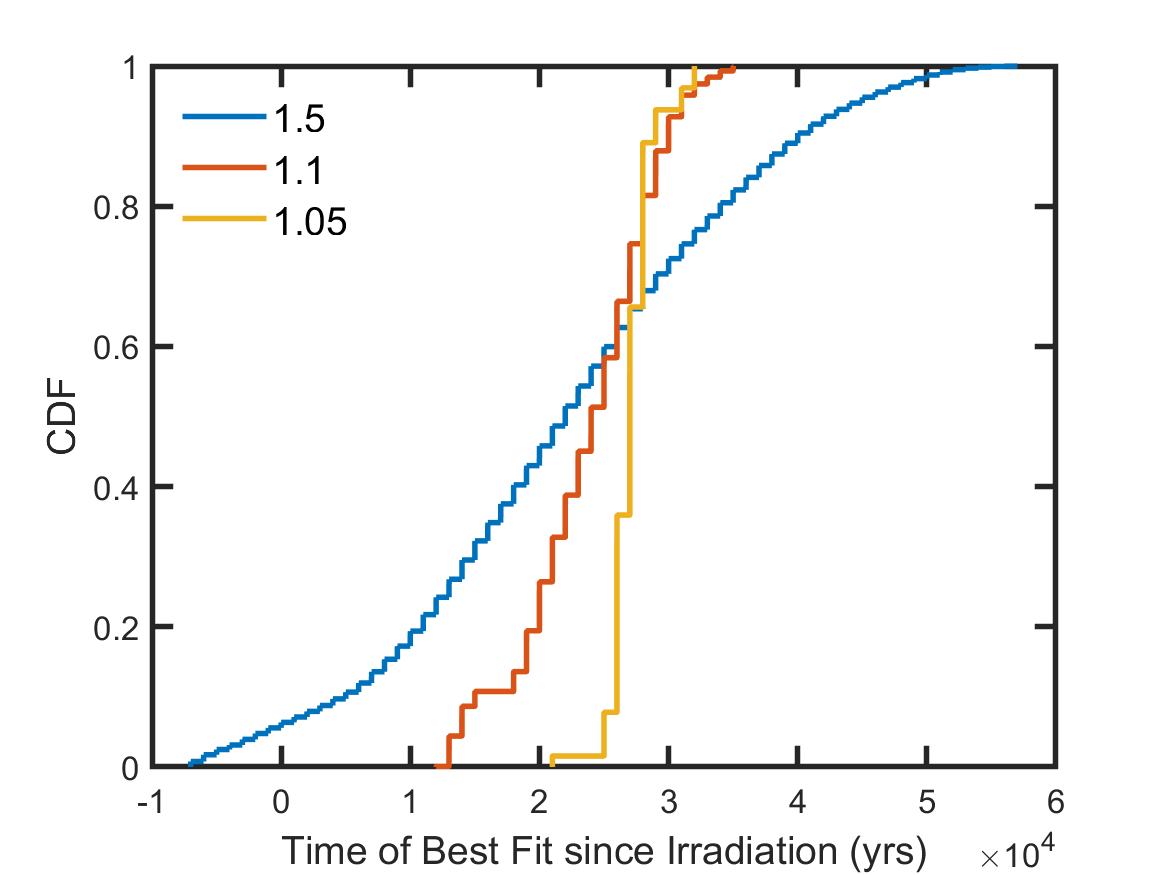}
\caption{Cumulative distribution function showing the time in a disc lifetime that best fits observed values. Time since irradiation is determined relative to the shielding time. Blue lines shows cumulative distribution functions for best fit instances with differences of up to factor 1.5, whilst red shows for differences up to factor 1.1, and yellow lines show for differences of up to factor 1.05.}
\label{fig:best_times}
\end{figure}

\subsubsection{Effect of initial disc compactness}

With the accretion rate being a measured observable, it is important to explore the parameters that can affect how much material is able to accrete on to the stars. These include the $\alpha_{\rm DW}$ parameter, the disc mass, and the scale radius. Above, we determined the simulations of best fit for different values of $\alpha_{\rm DW}$ and initial disc mass, but both for a single scale radius of 50 $\au$. Figure \ref{fig:alpha_rc_wind} shows the contours of simulation parameters that best fit observations, with colours denoting different initial scale radii. The solid contours show simulations that matched observed values to within factor 1.1, whilst dashed-dotted contours show areas with differences of factor 1.5. The different panels differentiate between different shielding times of: 0 Myr (top-left), 0.1 Myr (top-right), 0.3 Myr (bottom-left) and 0.5 Myr (bottom-right). Looking at the top-left panel the green points show the same contours as those in Fig. \ref{fig:contour_wind_50_0}, where the scale radius was equal to 50$\au$. By reducing the scale radius, the simulated discs that best agreed with observations moved to much smaller disc masses and weaker values of $\alpha_{\rm DW}$. The reduction in $\alpha_{\rm DW}$ is due to the discs being more compact, and so more mass was situated at smaller radii nearer to the central star. With larger gas surface densities there, a weaker $\alpha_{\rm DW}$ is required to better match the observed accretion rates. Moving to longer shielding times, it was previously mentioned that the best fit parameters moved to larger disc masses and values of $\alpha_{\rm DW}$, for a single value of the scale radius, 50 $\au$. Looking at the different panels of Fig. \ref{fig:alpha_rc_wind}, this is also the case for the evolution of the discs with different scale radii, as more mass and larger $\alpha_{\rm DW}$ values were required for the discs to match observations when they became unshielded.

\subsubsection{Determination of a disc irradiation age}

A common outcome of all of the simulations is that the time at which the simulations best fit the observations was always shortly after the discs became exposed to the high FUV irradiation. For discs with no shielding, this occurred very early in their lifetime, after 0.03 Myr as shown in Fig. \ref{fig:example_best}. However, for discs with some level of shielding, the time of best fit was later in their lifetimes, but always shortly after their shielding time. For example, for discs with shielding time of 0.5 Myr that could match the observed values, the time of best fit was after $\sim0.52$ Myr, therefore 0.02 Myr after the disc became irradiated. This was the case for all of the discs that could more precisely match the observations. Figure \ref{fig:best_times} shows the time in the simulations where the discs could best match observed values. The time is taken relative to the shielding time, i.e. it is the time of irradiation. The blue lines shows the discs that can match the observations to within a factor 1.5, whilst the red and yellow lines show discs that match observations to within factor 1.1 and 1.05 respectively. As can be seen by the yellow line shows those discs that best match the observations, the times at which they best fit was between 0.021 and 0.032 Myr. Even for those discs that match the observations to within factor 1.1 (red line), the times of best fit were between 0.013 and 0.035 Myr. The times of best fit shown in Fig. \ref{fig:best_times} indicate that d203-504 is either very young, or has just moved into the region of intense radiation, i.e. just emerged from the Orion Bar. This is in agreement with recent observations of d203-504 that find it's C/O ratio in the inner disc region is close to Solar indicating that the disc is consistent with that of a young, isolated, viscously evolving disc around a Solar-type star \citep{Schroetter25}.

\subsection{Viscous discs}
\label{sec:viscous}
\begin{figure*}
\centering
\includegraphics[scale=0.55]{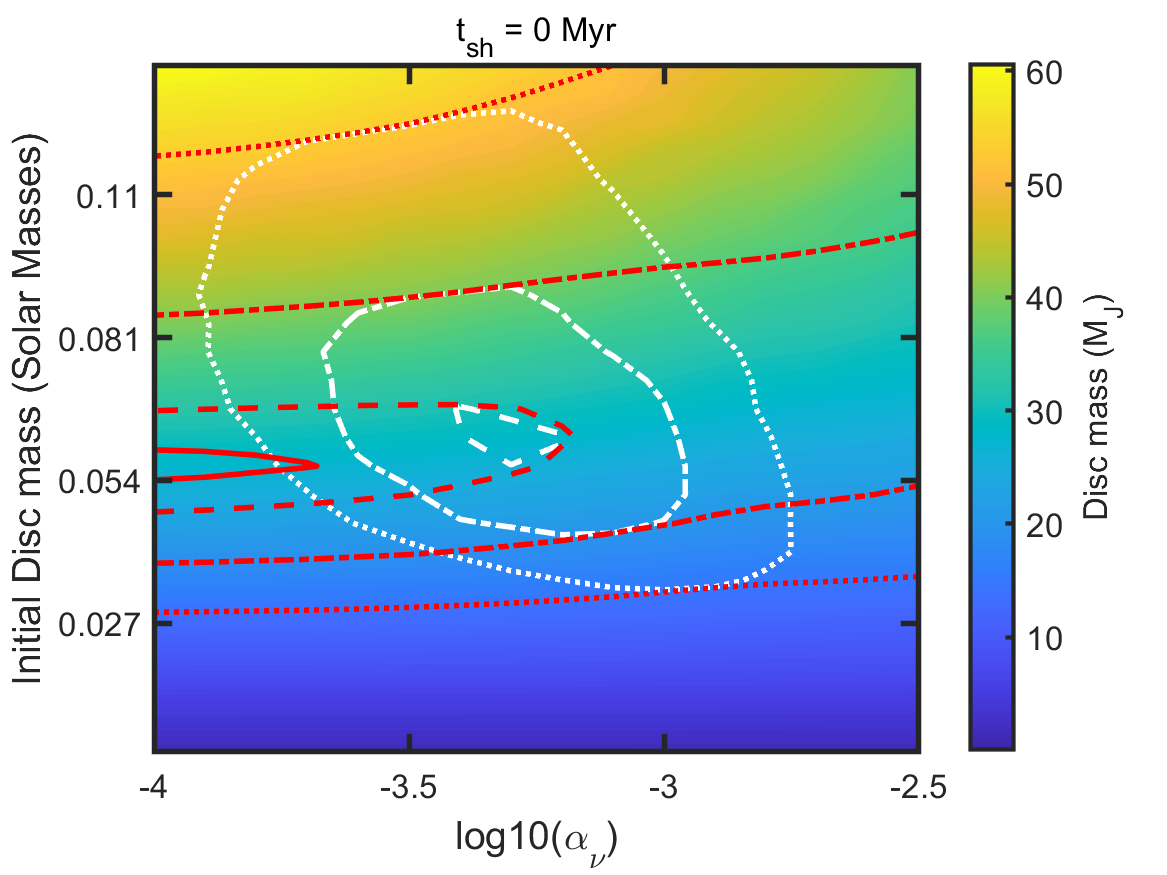}
\includegraphics[scale=0.55]{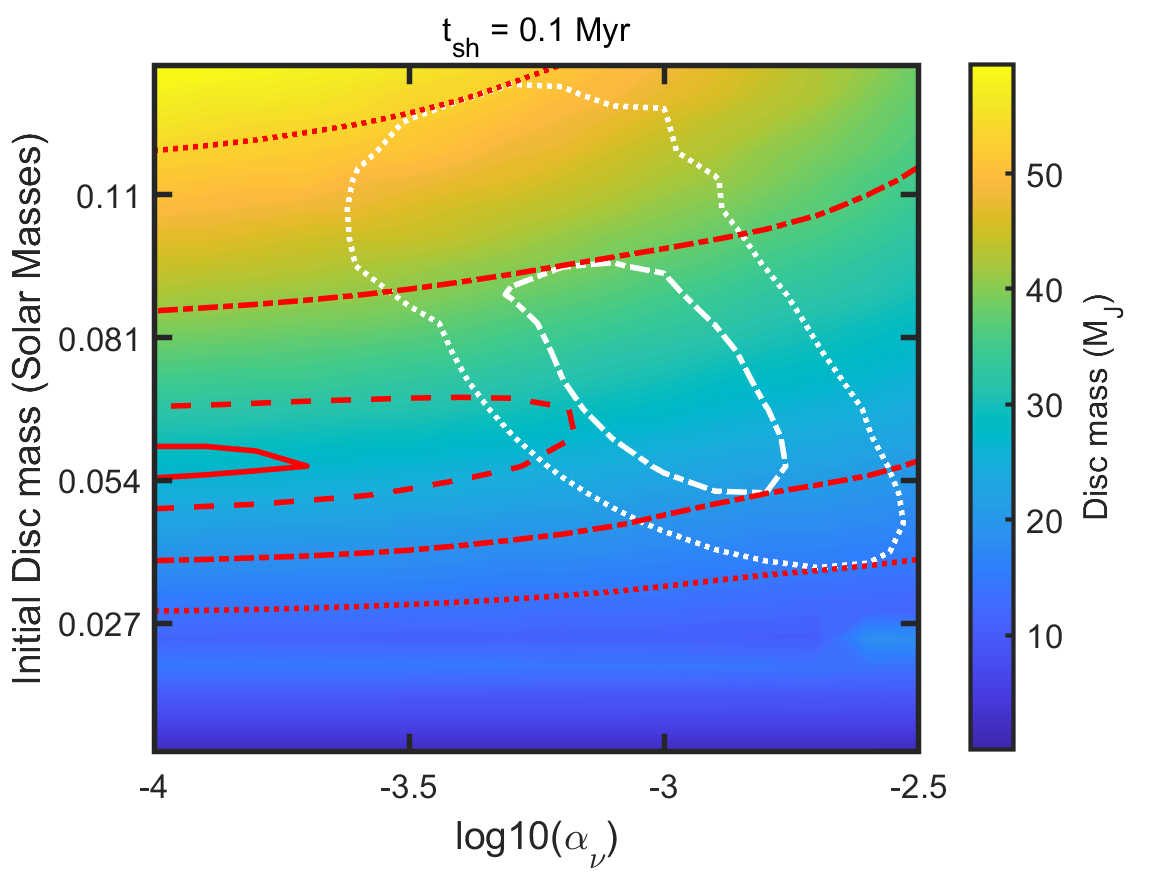}\\
\includegraphics[scale=0.55]{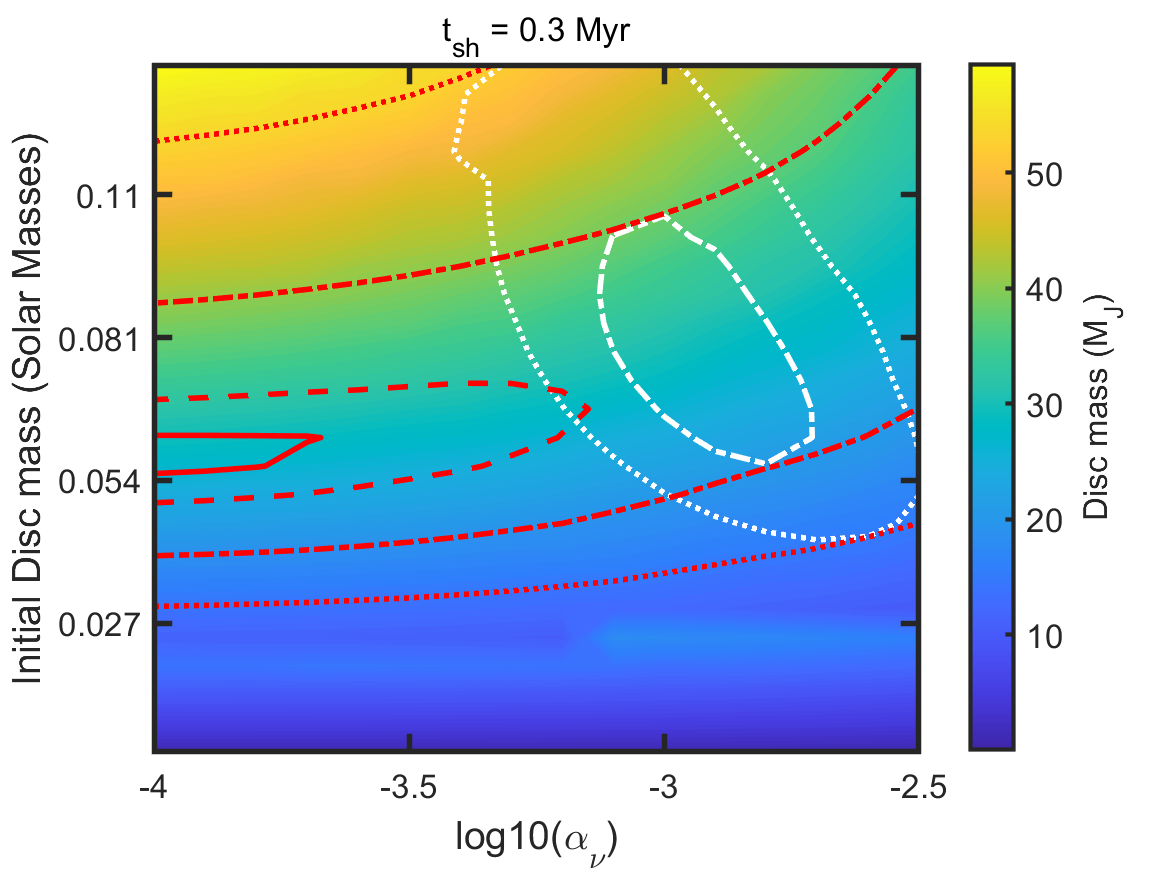}
\includegraphics[scale=0.55]{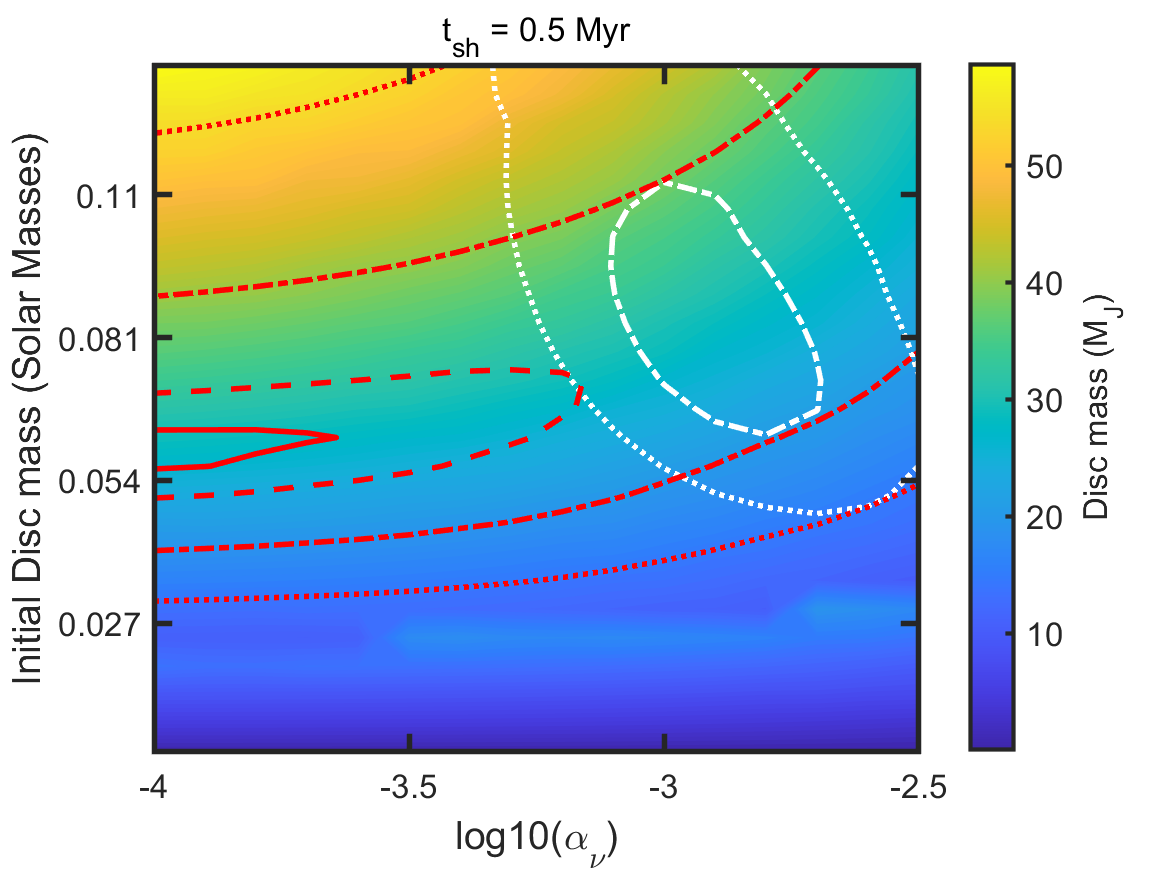}
\caption{Contour plot showing the disc mass for viscous discs at the time of best fit to the observed data for d203-504. The white contours highlight regions where the differences in log space between the simulations and observations are less than 0.041 (solid), 0.079 (dashed), 0.176 (dot-dashed) and 0.3 (dotted). These values correspond to differences of factors: 1.1, 1.2, 1.5 and 2 respectively. The red lines shows the same as the white lines, except the observed accretion rates are not compared with the simulations, highlighting the importance of knowing the accretion rate for determining compatible values of $\alpha$. The different panels denote different shielding times of: no shielding (top-left panel), 0.1 Myr (top-right panel), 0.3 Myr (bottom-left panel), and 0.5 Myr (bottom right panel).}
\label{fig:contour_visc_all}
\end{figure*}

Whilst Sect. \ref{sec:viscous} explored whether the observed values could be matched by MHD wind driven discs, we now explore whether viscous discs are equally able to explain the observations. In Fig. \ref{fig:contour_visc_all} we show the same contour plots as those in Fig. \ref{fig:contour_wind_all} highlighting the areas of parameter space where the simulated discs are best able to match observations. The different panels denote different shielding times of: no shielding (top-left panel), 0.1 Myr (top-right panel), 0.3 Myr (bottom-left panel), and 0.5 Myr (bottom right panel). Similar to Fig. \ref{fig:contour_wind_all}, the best fit regions are clearly evident by the white contours in the middle of each parameter. In terms of the initial disc mass, the discs that best fit the observations required initial disc masses between 0.05--0.09$\msun$, similar to that found for the MHD wind driven discs. However in regards to $\alpha_{\rm \nu}$, there is a wider spread of values that can adequately match observations to within factor 1.5 ranging between $10^{-3.5}$--$10^{-2.7}$. Interestingly, the viscous discs here could not match the observations as well as the MHD wind driven discs, shown by the lack of solid contours, whilst the required $\alpha_{\rm \nu}$ values are also at the upper end and beyond those expected from observations of turbulence in protoplanetary discs \citep{Lesur23}.

\begin{figure*}
\centering
\includegraphics[scale=0.55]{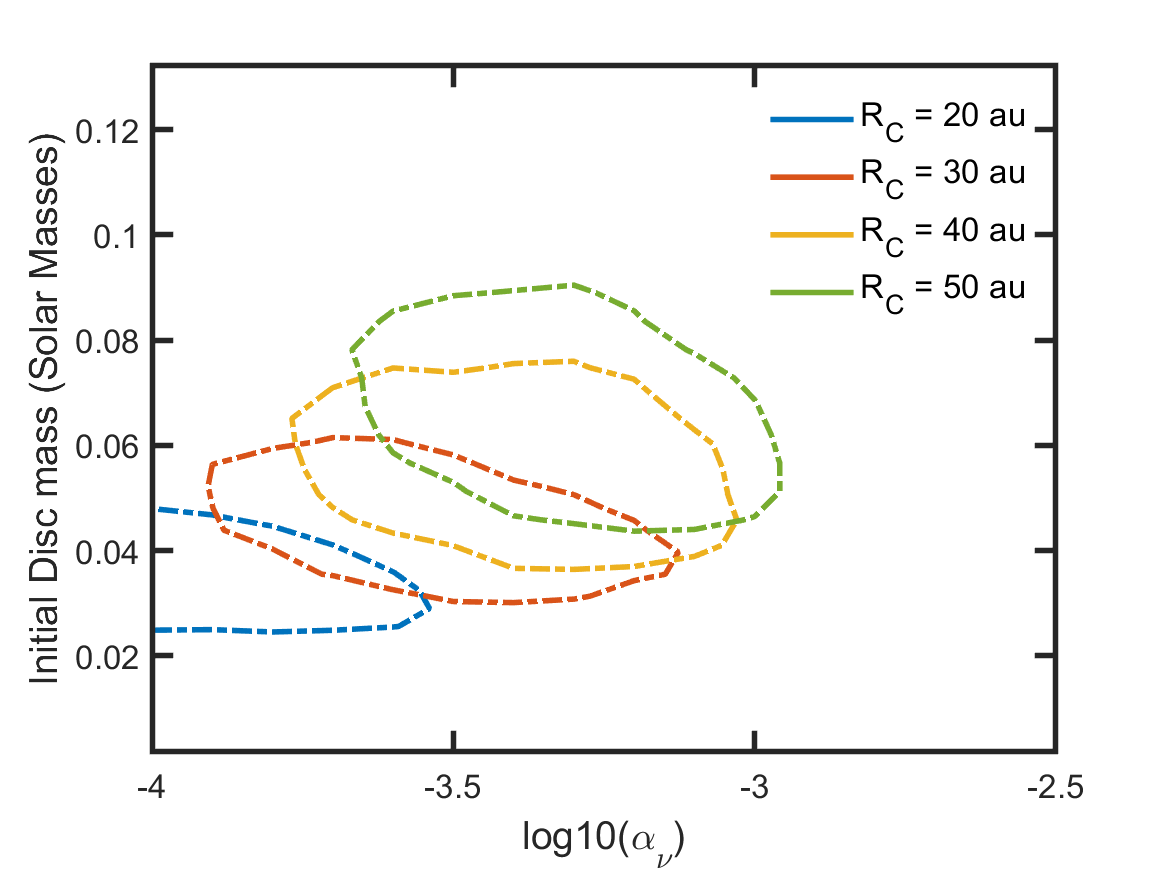}
\includegraphics[scale=0.55]{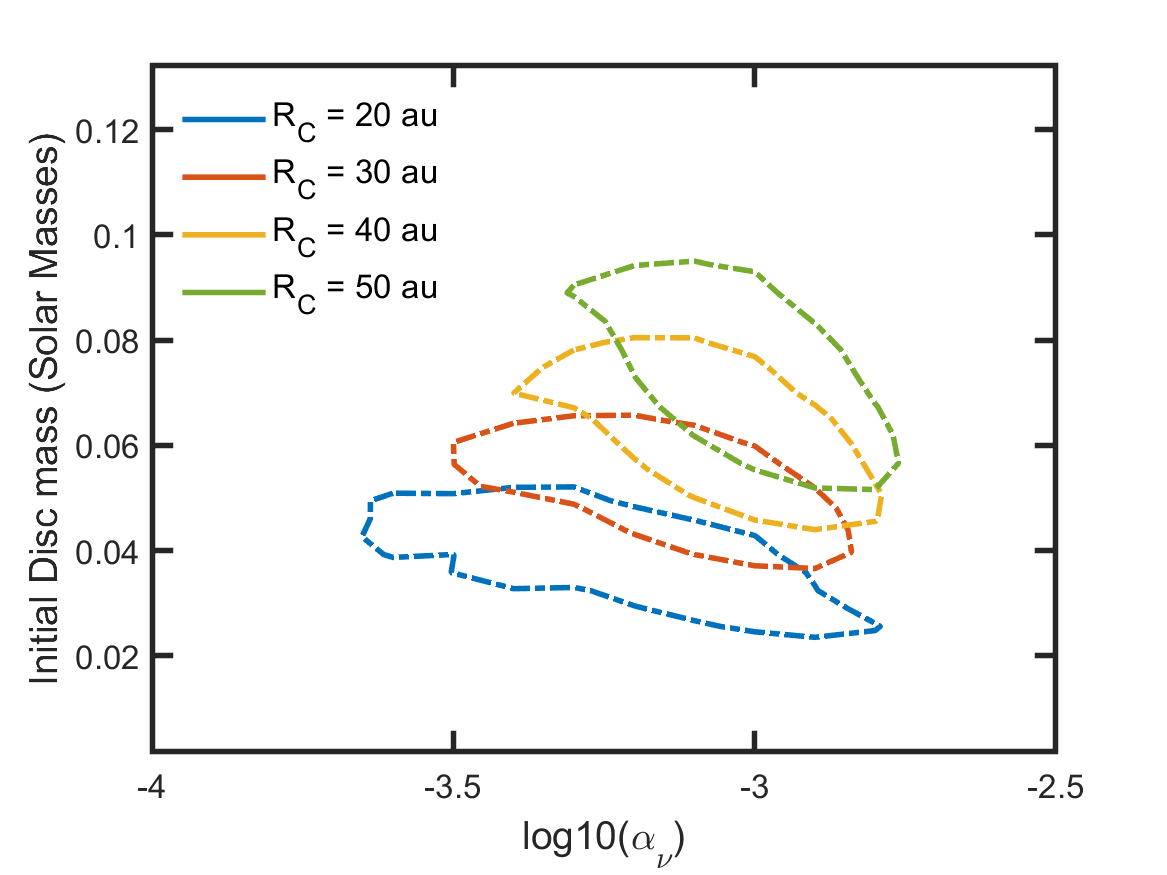}\\
\includegraphics[scale=0.55]{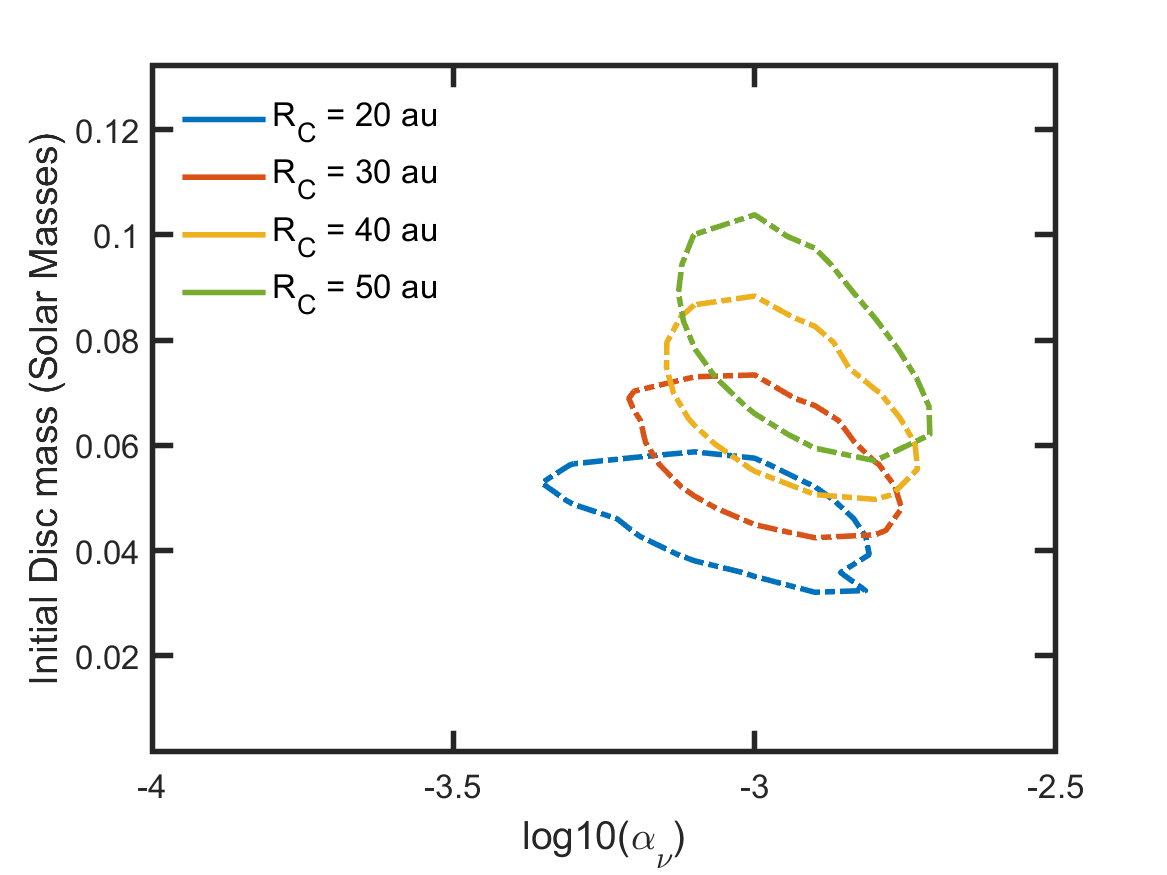}
\includegraphics[scale=0.55]{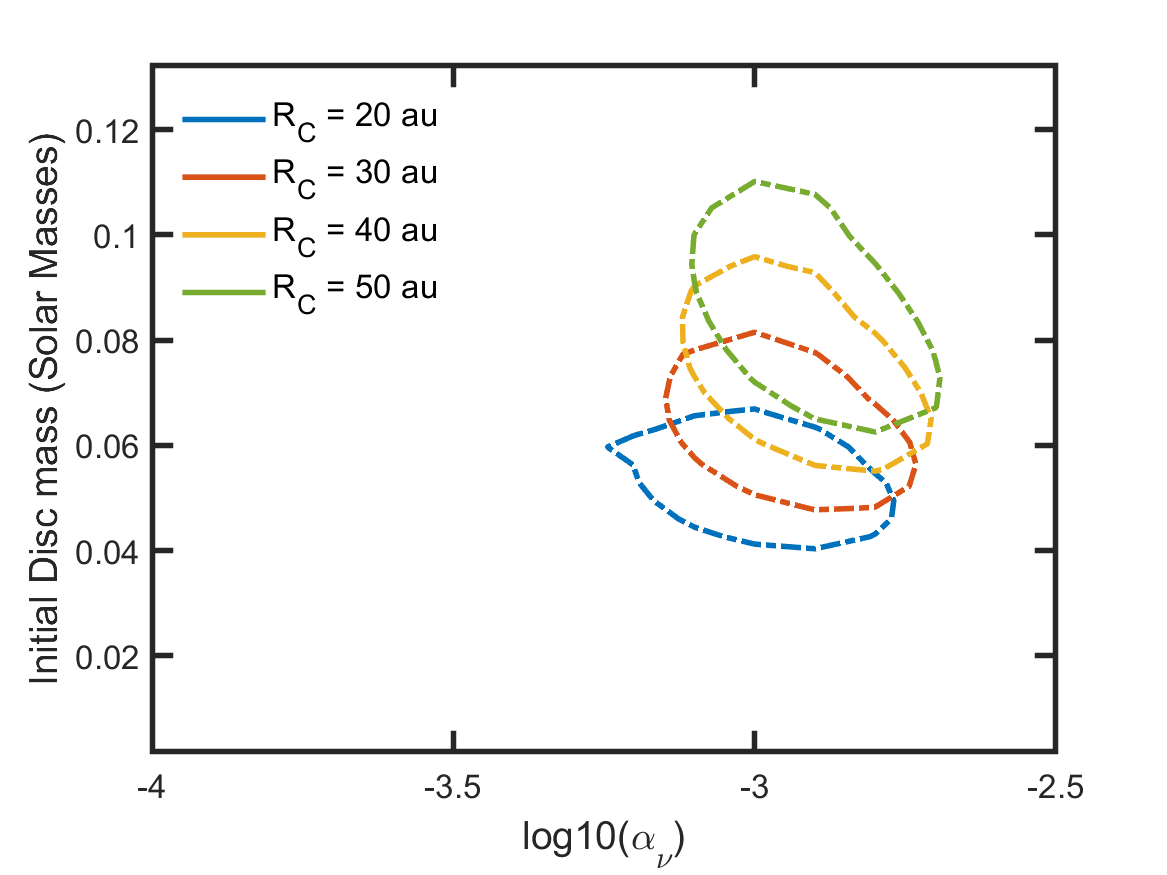}
\caption{Same as Fig. \ref{fig:alpha_rc_wind} but for viscous discs instead of MHD wind driven discs.}
\label{fig:alpha_rc_visc}
\end{figure*}

With Fig. \ref{fig:contour_visc_all} showing that viscous discs require similar initial masses to MHD wind driven discs, it only focussed on discs with scale radii of 50$\au$. Similar to MHD wind driven discs, initially more compact viscous discs also require lesser initial disc masses to sufficiently match the observed values for d203-504. Figure \ref{fig:alpha_rc_visc} shows the contours of simulation parameters that best fit observations for viscous discs, with colours denoting different initial scale radii, ranging from 20$\au$ (blue) to 50$\au$ (green). Indeed Fig. \ref{fig:alpha_rc_visc} shows that the most compact discs, $\sim20\au$, require initial disc masses of 0.02--0.07$\msun$, whilst the most extended discs, $\sim50\au$, required disc masses of 0.04--0.12$\msun$, similar to that required for MHD wind driven discs in order to match observations. The range in disc masses required to match observations highlights the impact of the shielding time. The smaller initial disc masses correspond to when there is little to no shielding time, whilst the more massive discs can be found for longer shielding times as those discs need to remain more massive for longer in order to match the observed accretion and photoevaporation rates.

Similar to the MHD wind driven discs, in order for the viscous discs to match the observed values for d203-504, they either needed to be young or they had to have just appeared from a shielded environment. For the viscous discs, their irradiation ages needed to be between 0.019--0.033 Myr, nearly identical to that found for MHD wind driven discs. This agreement in irradiation times is due to the dominance of external photoevaporation in the outer disc regions, that determines the agreement between the simulations and the observed external photoevaporation rates. With there being little difference between the irradiations times, this is also shows that there is little difference in the evolution of the outer regions of protoplanetary discs when using either viscous or MHD wind accretion scenarios, consistent with that seen in previous works \citep{Coleman24MHD}.

\section{Discussion}
\label{sec:constraints}

\subsection{Putting constraints on disc evolution parameters}

With Sect. \ref{sec:results} showing that both viscous and MHD wind driven discs were able to match the observables found for d203-504, we now compare the constraints that the simulations place on disc parameters with those expected from other observations and theoretical expectations. From observations, there are multiple avenues for constraining the strength of $\alpha$, typically by comparing observations of protoplanetary discs with theoretical models to see what parameters are necessary for good agreement. For example, the vertical extent of pebbles and dust is set by the strength of vertical turbulence in the disc, where recent observations have shown that $\alpha_{\nu}\leq 10^{-4}$ \citep{Villenave22}. Additionally, like the vertical extent of dust discs, the radial extent of dust structures compared to gas structures is also dependent on the level of turbulence, since more turbulent discs will act to smear out dust features compared to those with weak turbulence. Through ALMA, numerous features in multiple discs have been observed constraining $\alpha_{\nu}$ to between $10^{-4}$--$10^{-2}$ \citep{Dullemond18,Sierra19,Rosotti20}. The radial extent of the disc, and its subsequent evolution, can also be used as a measure of turbulence, since either discs are born large and constantly truncate, or they need to reach an equilibrium where outward transport is required to supply the mass lost through disc winds. Observations of populations of discs have placed upper limits of $\alpha_{\nu}\leq 10^{-3}$ since few large discs are observed, indicating a moderate level of equilibrium with the local environment \citep{Toci21}.

Asides from the radii of discs or substructures, and their properties, the level of accretion from the disc on to the central star has long been used as a general measure of turbulence. With the accretion rates depending not only on the the rate of angular momentum transport through the disc, of which $\alpha$ is a proxy for, but also on the disc mass, this creates a degeneracy between the disc mass and $\alpha$. Note here that $\alpha$ isn't representative of turbulence, and more of the rate of angular momentum transport through the disc and on to the central star, and so may be a better measure of the underlying $\alpha_{\nu}$ or $\alpha_{\rm DW}$ in protoplanetary disc models. Nonetheless, multiple works have utilised populations of discs with known mass accretion rates to place constraints on $\alpha$, showing that there exists a wide range with \citet{Ansdell18} obtaining values between $10^{-4}$--$10^{-1}$, whilst \citet{Trapman20} find $\alpha\sim 10^{-5}$--$10^{-2}$.

\begin{figure*}
\centering
\includegraphics[scale=0.55]{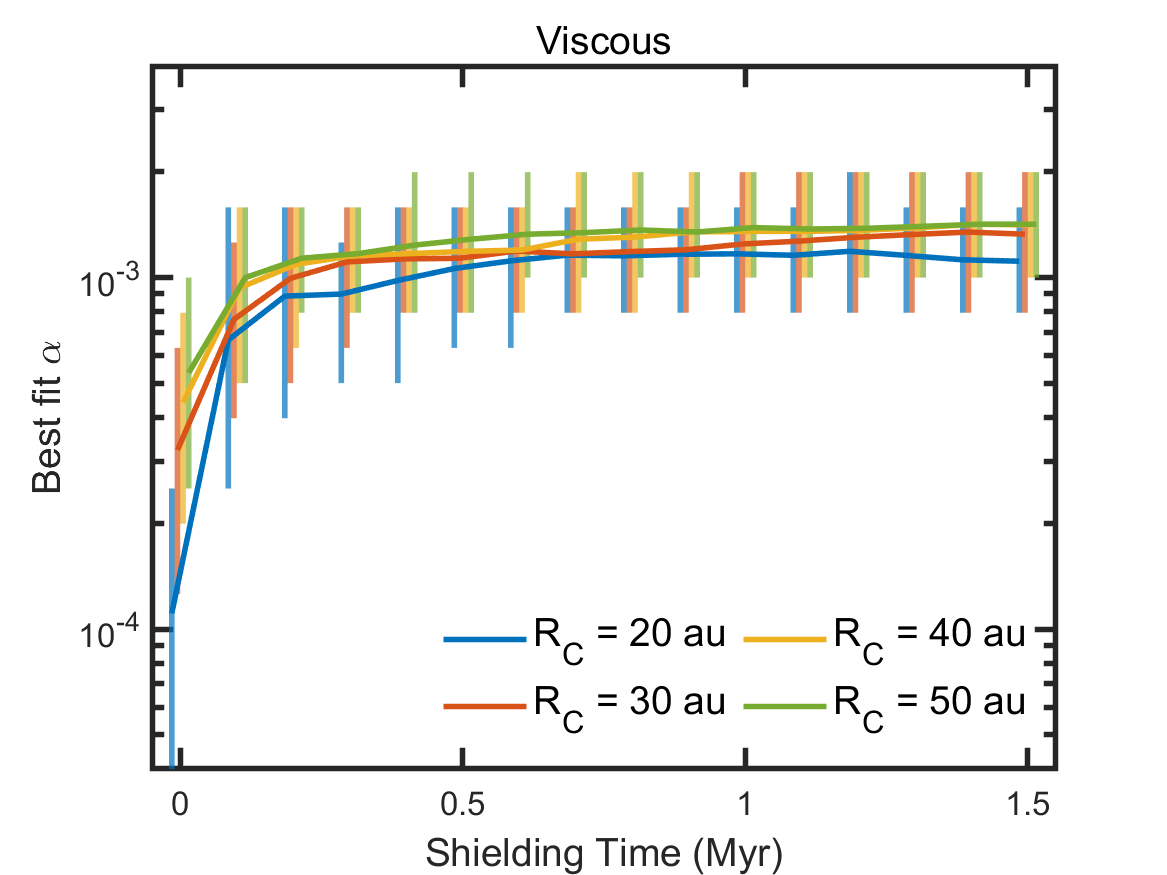}
\includegraphics[scale=0.55]{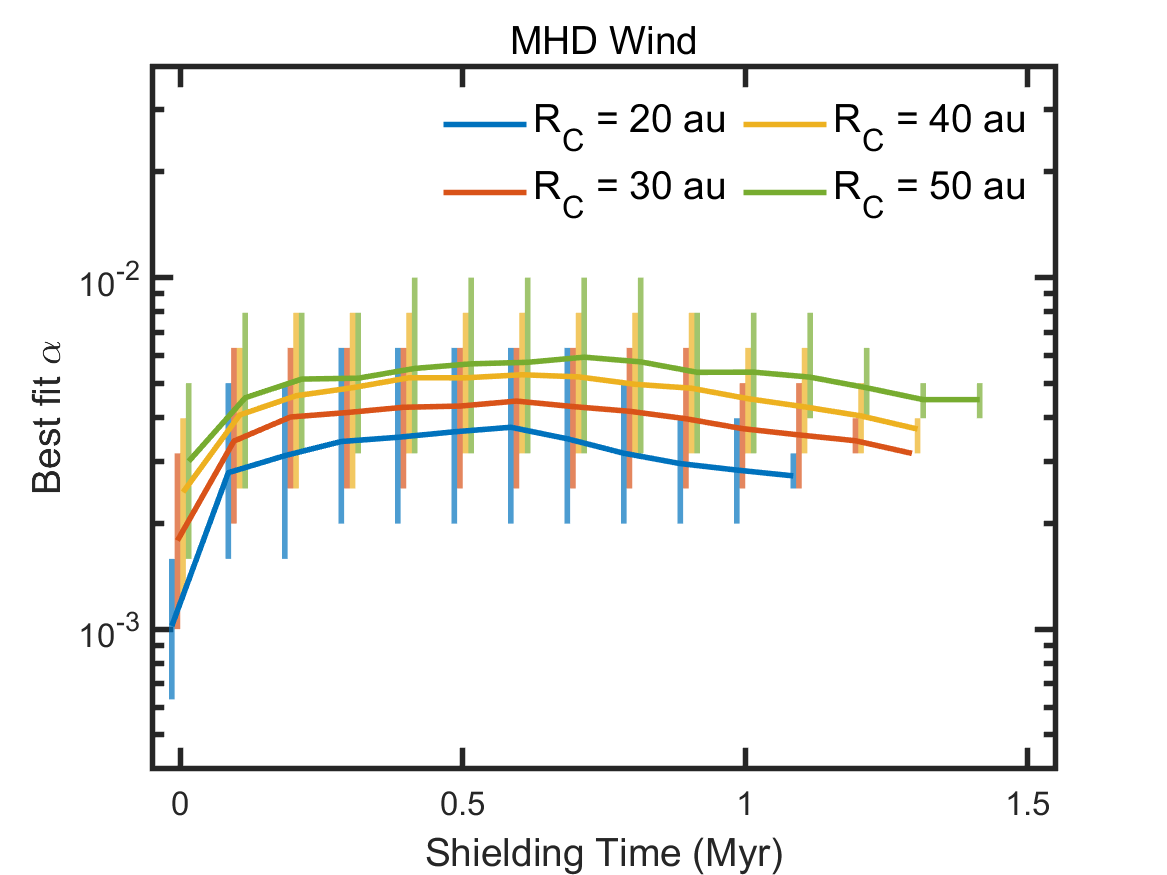}
\caption{The best fit values for $\alpha$ taken as those with differences of up to factor 1.5, from viscous (left-hand panel) and MHD wind driven (right-hand panel) discs as a function of the shielding time. The different colour segments show the changes in the best fit $\alpha$ values as a function of the initial compactness of the discs.}
\label{fig:best_alpha}
\end{figure*}

From theoretical standpoints, there are multiple processes that produce turbulence in protoplanetary discs that have been studied in recent years. These include both thermo-hydrodynamic and magneto-hydrodynamic instabilities. For the thermo-hydrodynamic instabilities, these operate in the midplane and lower layers of disc, where it is usually optically thick and weakly ionised. The main instabilities that operate there are the vertical shear instability (VSI, \citet{Arlt04,Nelson13}), the convective overstability (COS, \citet{Klahr14,Lyra14}), and the zombie vortex instability (ZVI, \citet{Barranco05}). The VSI, powered by the differential rotation in the disc's vertical structure generates turbulence in the outer regions of protoplanetary discs with an $\alpha_{\nu}\simeq 10^{-4}$ \citep{Flock17,Flock20}. Meanwhile, the convective overstability instability arises where radially buoyant gas performs radial oscillations on epicyclic frequencies. This depends on the cooling time-scale of the gas, compared to the orbital period, as the gas can exchange heat with the surrounding material, experiencing a buoyant acceleration which causes the oscillations that can eventually go unstable and produce large-scale vortices \citep{Klahr14,Lyra14,Latter16}. These instabilities have been found to generate turbulence up to $\alpha_{\nu}\sim10^{-3}$, but this can vary by an order of magnitude depending on the the buoyancy and cooling parameters \citep{Raettig13}. The final instability, the Zombie vortex instability, is a non-linear instability that can develop if there is sufficient vertical buoyancy. This can generate long-lived vortices in the midplane of planetary discs, where there is little underlying turbulence, i.e. a dead zone \citep{Barranco05}. These vortices then interact with the surrounding disc, generating turbulence equivalent to $\alpha_{\nu}\simeq10^{-5}$--$10^{-4}$ \citep{Barranco05,Barranco18}.
Ultimately, from theoretical studies, instabilities can generally result in turbulence on the order $\alpha_{\nu}\simeq 10^{-5}$--$10^{-3}$.

With instabilities being the main mechanisms for generating turbulence in protoplanetary discs, it is the level of magnetization that determines the strength of the MHD disc winds. it has been found that $\alpha_{\rm DW}$ is roughly proportional to the level of magnetization of the disc, that is the inverse of the ratio of the thermal-to-magnetic pressure in the disc mid-plane \citep{Bai13,Bai13b}. Indeed \citet{Bai13b} found torques equivalent to $\alpha_{\rm DW}=10^{-5}$--$10^{-3}$ with local MHD simulations. This corresponds to magnetization levels of the disc $\beta_0=10^3$--$10^5$, where $\alpha_{\rm DW}\propto \beta_0^{-1}$.

With observations of local disc properties indicating $\alpha_{\nu}\simeq 10^{-4}$--$10^{-2}$ and accretion rates implying $\alpha\simeq 10^{-5}$--$10^{-1}$, it is still uncertain as to what the strength $\alpha$ is in protoplanetary discs. 
Additionally, theoretical processes are mostly only able to generate turbulence equivalent $\alpha\leq10^{-4}$, except for the convective overstability that can generate turbulence $\alpha\leq10^{-3}$, whilst local MHD simulations indicate $\alpha_{\rm DW}=10^{-5}$--$10^{-3}$ resulting from the level of magnetization in the discs.
The simulation results presented in Sect. \ref{sec:results} equally showed a range of $\alpha$ values that could adequately match observations. In Fig. \ref{fig:best_alpha}, we show the best fit $\alpha$ values as a function of the shielding time for viscous (left panel) and MHD wind driven (right panel) discs. We take the best fit values as those that are from discs with differences of up to a factor 1.5 to the observations. The colours denote different initial scale radii, representing the initial compactness of the discs. The trends in how the best fit $\alpha$ value changes as a function of the shielding time is similar for both viscous and MHD wind driven discs. For discs with no shielding time, i.e. the discs had only just formed, then the best fit values for $\alpha$ are between $10^{-4}$--$10^{-3}$ for viscous discs, and $10^{-3}$--$7\times10^{-3}$ for MHD wind driven discs. The variance for both viscous and MHD wind driven discs arises due to the initial disc mass and scale radius, where more massive discs required weaker values of $\alpha$, and more extended discs required larger values of $\alpha$. As the shielding time increases, the best fit $\alpha$ values increase initially before settling on a consistent range independent of the shielding time. For viscous discs, the best fit $\alpha$ settled between $4\times 10^{-4}$--$2\times 10^{-3}$, whilst for MHD wind driven discs, the best fit $\alpha$ values settled between $2\times 10^{-3}$--$10^{-2}$. Similar to the case with no shielding, the initial compactness and mass of the disc provides the spread in the best fit values for $\alpha$, where additional constraints on those parameters would further constrain the fits to $\alpha$ from the simulations. We also note that changing the magnetic lever arm value to between 2--6, as is shown in Appendix \ref{appendix:lever}, also gives a range of compatible values for $\alpha$ whilst converging to smaller values as the lever arm is increased.

In comparing our best fit $\alpha$ values to observations, the levels for the viscous $\alpha_{\nu}$, are compatible with those for accretion on to the central stars \citep{Ansdell18,Trapman20}, as well as the radial extent of protoplanetary discs \citep{Toci21}, and the local properties of substructures \citep{Dullemond18,Sierra19,Rosotti20}. Whilst the values for $\alpha$ are compatible with those derived from studies of accretion rates, they are towards the lower end of those estimates, and so may struggle to explain some of the larger values in those studies. However, whilst they agree with some of the observations, albeit to the upper end of observations, like the observations, the best fit $\alpha_{\nu}$ values are larger than those that can be derived from theoretical models \citep[e.g.][]{Nelson13,Klahr14,Barranco05}. For the MHD wind driven cases, the best fit $\alpha$ values can only be compared with the surveys of accretion rates on to the central stars, since MHD winds are not responsible for turbulence in the disc that can determine the shape and properties of substructures as well as the vertical extent of dust and pebble layers. When comparing with the accretion rates from observation studies, the best fit $\alpha$ values are in good agreement with those from earlier \citep{Hartmann98,Andrews09,Andrews10} and more recent studies \citep{Ansdell18,Trapman20}, indicating higher levels of $\alpha$ than typically considered for viscous disc models, i.e. $\alpha\ge10^{-3}$, which would allow the MHD wind driven models to better explain the larger observed values of $\alpha$ for individual discs.

More recently, a number of works using MHD wind driven models have attempted to place constraints on the value for $\alpha_{\rm DW}$ in order to match disc lifetimes of $\sim$few Myr \citep{Tabone22,Tabone25}. In those works they find that $\alpha_{\rm DW}\leq10^{-3}$, lower than the values that we find are required to match the multitude of observational constraints  for d203-504. Whilst the two estimations for $\alpha_{\rm DW}$ may be contradictory, it is worth noting that \citet{Tabone22,Tabone25} consider discs with initial disc radii only up to the lower end of the parameter space explored here and for which we find the weakest viable solutions for d203-504. Applying a broader range of possible initial disc radii coupled with other dispersal processes such as internal and external photoevaporative winds may result in more consistent estimates of $\alpha_{\rm DW}$ between regions \citep[external photoevaporation has been shown to be important even in relatively weak UV environments as part of the ALMA AGE-PRO progamme,][]{Anania25}. However we also note that there is no strict requirement that $\alpha_{\rm DW}$ take on a universal value and so should differences in $\alpha_{\rm DW}$ persist they may indeed be genuine.

\subsection{The PAH-to-gas ratio of d203-504}
\label{sec:PAHtogas}
We have shown that the constraints on the combined physical parameters and dispersal rates of d203-504 tightly constrain models of the evolution of the system. However, we have utilised a fiducial external photoevaporation model with a PAH-to-gas ratio of 1/100 and \citet{Schroetter25} place a new constraint on that parameter for d203-504 of closer to 1/16. Here we discuss the constraints and implications placed on the model by this estimate.

The \textsc{friedv2} grids that we use for the external photoevaporative mass loss rate define a PAH-to-dust ratio and a dust-to-gas ratio \citep{Haworth23}. These combine to give an effective PAH-to-gas ratio. The \textsc{friedv2} grid has two modes for the dust: i) where the dust in the outer disc and hence the wind is ISM-like and ii) where grain growth and drift has occurred at the disc outer edge, depleting the dust entrained in the wind by a factor 100 \citep{Facchini16, 2025MNRAS.539.1414P}. The PAH-to-dust ratio is then set at either $f_{\rm PAH}=$0.1, 0.5 or 1 of the ISM value. Overall this means the options in \textsc{friedv2} for the PAH-to-gas ratio are 1 (ISM-like dust, $f_{\textrm{PAH}}=1$), 1/2 (ISM-like dust, $f_{\textrm{PAH}}=0.5$), 1/10 (ISM-like dust, $f_{\textrm{PAH}}=0.1$), 1/100 (dust depleted wind, $f_{\textrm{PAH}}=1$), 1/500 (dust depleted wind, $f_{\textrm{PAH}}=0.5$), 1/1000 (dust depleted wind, $f_{\textrm{PAH}}=0.1$).

\begin{figure}
    \centering
    \includegraphics[width=\linewidth]{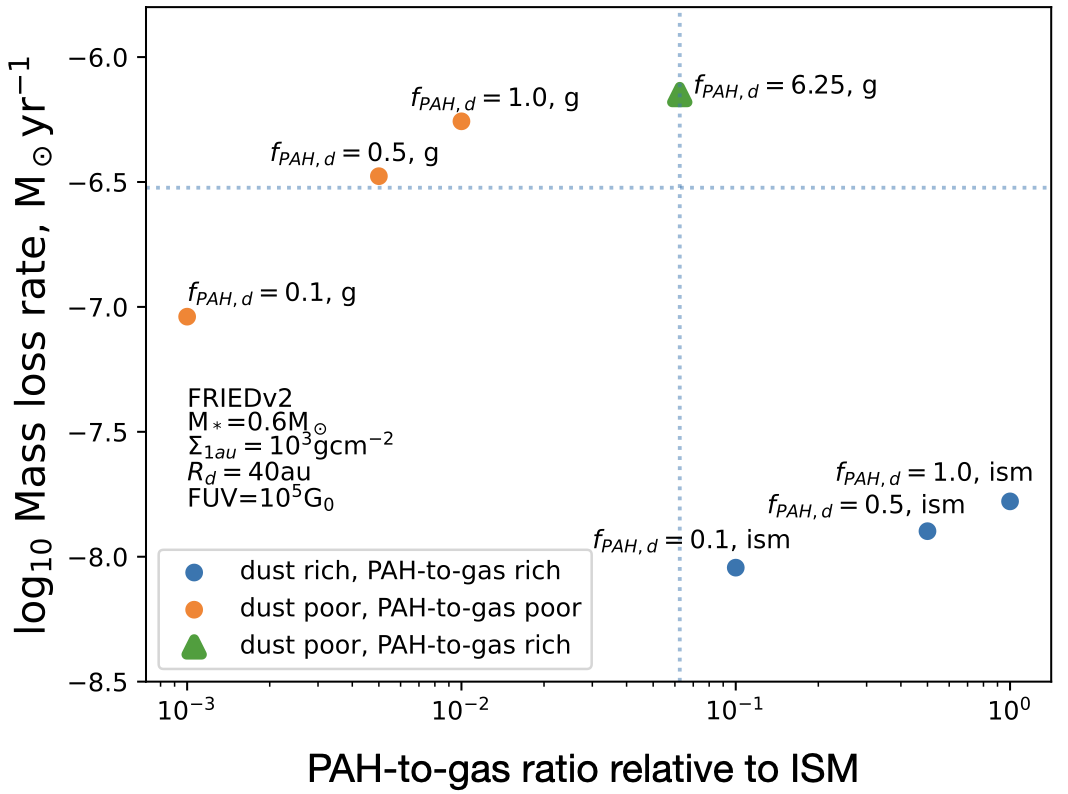}
    \caption{The mass loss rate as a function of PAH-to-gas ratio relative to the ISM for F\textsc{friedv2} external photoevaporation calculations. \textsc{FRIEDv2} samples values of the PAH-to-dust ratio $f_{\textrm{PAH},d}$ and has two options for either ISM-like dust (models in the lower right) or depleted dust in the wind due to grain growth and radial drift (models in the upper left). However it is possible to also have a dust depleted wind with high PAH-to-gas ratio (the triangle point) which would be more compatible with recent observational constraints for 203-504 and which does not give significantly different mass loss rates to the fiducial \textsc{friedv2} grid subset ($f_{\textrm{PAH},d}=1,g$ ) that we used in this paper. The dotted lines show the observed constraints on d203-504 for reference, but which we are not trying to fit with these mass loss rate calculations. }
    \label{fig:microphysicsMdot}
\end{figure}
From the above, there is no clear analogue within \textsc{friedv2} for the observed PAH-to-gas ratio for d203-504, and the only model that exists that gets close is one where there is ISM-like dust in the wind and a low PAH-to-dust ratio. We find that that subset of the \textsc{friedv2} grid, where small dust shields the disc and the PAH depletion means less heating, is incapable of producing sufficient mass loss to match the observed mass loss rate. We note that this discrepancy is at a level that cannot be compensated for through uncertainty e.g. in the disc mass, since the mass loss rate sensitivity to mass is quite weak \citep{Haworth23}.

A more natural solution would be that the dust rich wind with low PAH-to-dust ratio is not uniquely the way to obtain PAH-to-gas ratios similar to those observed. An alternative could be to have a dust-depleted wind where the PAH abundance has not been scaled down with the dust, i.e. where the PAH-to-dust ratio is larger than the fiducial value in the ISM. We find that in that case mass loss rates can be sufficiently high to match observations while maintaining a less depleted PAH-to-gas ratio. This is illustrated in Figure \ref{fig:microphysicsMdot}, which shows mass loss rates as a function of PAH-to-gas ratio relative to the ISM for a 0.6\,M$_\odot$ star with a 40\,au disc, surface density at 1\,au of $10^3$g\,cm$^{-2}$ and $10^5$\,G$_0$ radiation field. The triangular point is the new model with the enhanced PAH-to-dust ratio. So our models would require this latter scenario of a dust depleted, PAH rich wind to explain the high PAH-to-gas ratio and high mass loss rate for d203-504. However, further observations are required to prove this prediction.

We also note that the PAH-to-gas ratio derived from JWST observations remains highly uncertain at this stage, as it depends on the assumption that the total gas column in the PDR - where  PAH emission occurs -  is $5.5\times10^{21}$cm$^{-2}$ \citep{Schroetter25}. This value was originally estimated for the proplyd HST 10 by  \cite{StorzerHollenbach1999}, but it may not be applicable to d203-504. Furthermore, the dust abundance within the photoevaporative wind is currently unknown. These uncertainties clearly highlight the need for dedicated studies of dust and PAH properties in irradiated protoplanetary discs. JWST spectroscopic surveys of irradiated discs from Cycles 3 to 5, combined with existing ALMA data, will be instrumental in placing tighter constraints on the dust properties in these environments

Overall, this demonstrates that the combination of observational constraints used in this paper deepens our understanding of both the disc evolution and the wind microphysics. This will be further enhanced by applying similar analysis to a statistical sample in Orion. 

\section{Conclusions}
\label{sec:conc}

Recent observations of the proplyd d203-504 have provided estimates for the disc mass, size, mass accretion rate, and external photoevaporation. This has made d203-504 the first such proplyd with measured mass accretion and external mass loss rates. In this work we use protoplanetary disc evolution models to explore what parameters in those models can best fit the observed values. Our goals are to both develop a deeper  understanding of d203-504 itself and to motivate more generally how insightful having such a collection of empirical disc properties can be. We draw the following main conclusions from this work.\\

1. Both viscous and MHD wind driven disc models were able to simulate discs consistent with the observations. The simulated discs were able to simultaneously match the observed accretion rates, photoevaporative mass loss rates, and disc size and mass. Whilst both mechanisms could match the data, MHD wind driven discs did so to greater precision. Additionally, viscous and MHD wind driven discs preferred moderate initial disc masses between 0.05--0.09$\msun$, as well as discs that were initially more extended.\\

2. By including the effects of shielding the discs from external radiation, both viscous and MHD wind driven discs were able to match the observations at older disc ages. In order to match, those discs required larger initial disc masses, and either increased levels of turbulence or MHD wind strength. For both the viscous and MHD-wind driven discs, the simulated discs could match the observations, to varying degrees, for shielding times of up to 1.5 Myr. Accurately determining the age of the system would therefore significantly constrain the historical evolution of the disc.\\

3. All of the discs that were able to match the observations did so either with ages <0.04 Myr, or within 0.04 Myr of their shielding time ending. This was mainly a result of simultaneously matching the disc radius and large external photoevaporation rate, whilst also maintaining sufficiently high accretion rates on to the central star. We define this age as the disc irradiation age, highlighting how long discs have resided in high energy environments. This irradiation age provides insight into the proplyd lifetime problem by indicating that those discs with large external photoevaporation rates have not experienced those high rates for a long period of time, showing that they are a short-lived phenomena for those discs. This is consistent with previous discussion of the proplyd lifetime problem \citep[e.g.][]{Henney99, Winter2019}.\\

4. With both viscous and MHD wind driven discs able to match the observations, our simulations can place constraints on the levels of turbulence or the MHD wind strength. For viscous discs, the simulations could best match observations for $\alpha_{\nu}$ values between $10^{-4}$--$10^{-3}$, whilst for MHD wind driven discs, the best matches were for $\alpha_{\rm DW}\sim 10^{-3}$--$7\times 10^{-3}$. These ranges however, were when there was no shielding time. When the discs experienced some level of shielding, the best fit values for $\alpha_{\nu}$ was between $3\times 10^{-4}$--$2\times 10^{-3}$ for viscous discs, and $\alpha_{\rm DW}\sim 2\times 10^{-3}$--$10^{-2}$ for MHD wind driven discs.\\

In summary, the simulations presented in this work are able to match the observed values for d203-504 found in \citet{Schroetter25}. They place constraints on the expected initial disc masses and sizes, as well as the strengths of either turbulence or MHD winds. When comparing the constraints on the values for $\alpha$ in particular with other works, the simulations here are favouring discs that are MHD wind driven, since the predictions for viscous discs indicate that $\alpha$ is between $5\times10^{-3}$--$2\times10^{-4}$ for discs that had some level of shielding. Such high values of $\alpha$ are towards the upper end of what is observed in levels of turbulence from vertical settling \citep{Villenave22}, as well as from theoretical processes. For MHD wind driven discs however, the expected values of $\alpha_{\rm DW}\sim 2\times 10^{-3}$--$10^{-2}$ are consistent with those required to match accretion rates on to central stars \citep{Ansdell17}. The constraints placed on disc parameters in this work are only based on a single disc, d203-504. Should future observations of other discs provide similar data, then they could be used to provide more accurate constraints on fundamental disc properties.

\section*{Data Availability}
The data underlying this article will be shared on reasonable request to the corresponding author.

\section*{Acknowledgements}
GALC acknowledges funding from the Royal Society under the Dorothy Hodgkin Fellowship of T. J. Haworth, and UKRI/STFC grant ST/X000931/1. TJH  acknowledges a Dorothy Hodgkin Fellowship, UKRI guaranteed funding for a Horizon Europe ERC consolidator grant (EP/Y024710/1) and and UKRI/STFC grant ST/X000931/1.
OB and IS are funded by the Centre Nationald’Etudes Spatiales (CNES) through the APR program. This research received funding from the program ANR-22-EXOR-0001 Origins of the Institut National des Sciences de l’Univers, CNRS.
This research utilised Queen Mary's Apocrita HPC facility, supported by QMUL Research-IT (http://doi.org/10.5281/zenodo.438045). This work used the DiRAC Data Intensive service (DIaL2 / DIaL [*]) at the University of Leicester, managed by the University of Leicester Research Computing Service on behalf of the STFC DiRAC HPC Facility (www.dirac.ac.uk). The DiRAC service at Leicester was funded by BEIS, UKRI and STFC capital funding and STFC operations grants. DiRAC is part of the UKRI Digital Research Infrastructure.

\bibliographystyle{mnras}
\bibliography{references}{}

\appendix
\section{Effects of changing magnetic lever arm $\lambda$}
\label{appendix:lever}
\begin{figure*}
\centering
\includegraphics[scale=0.55]{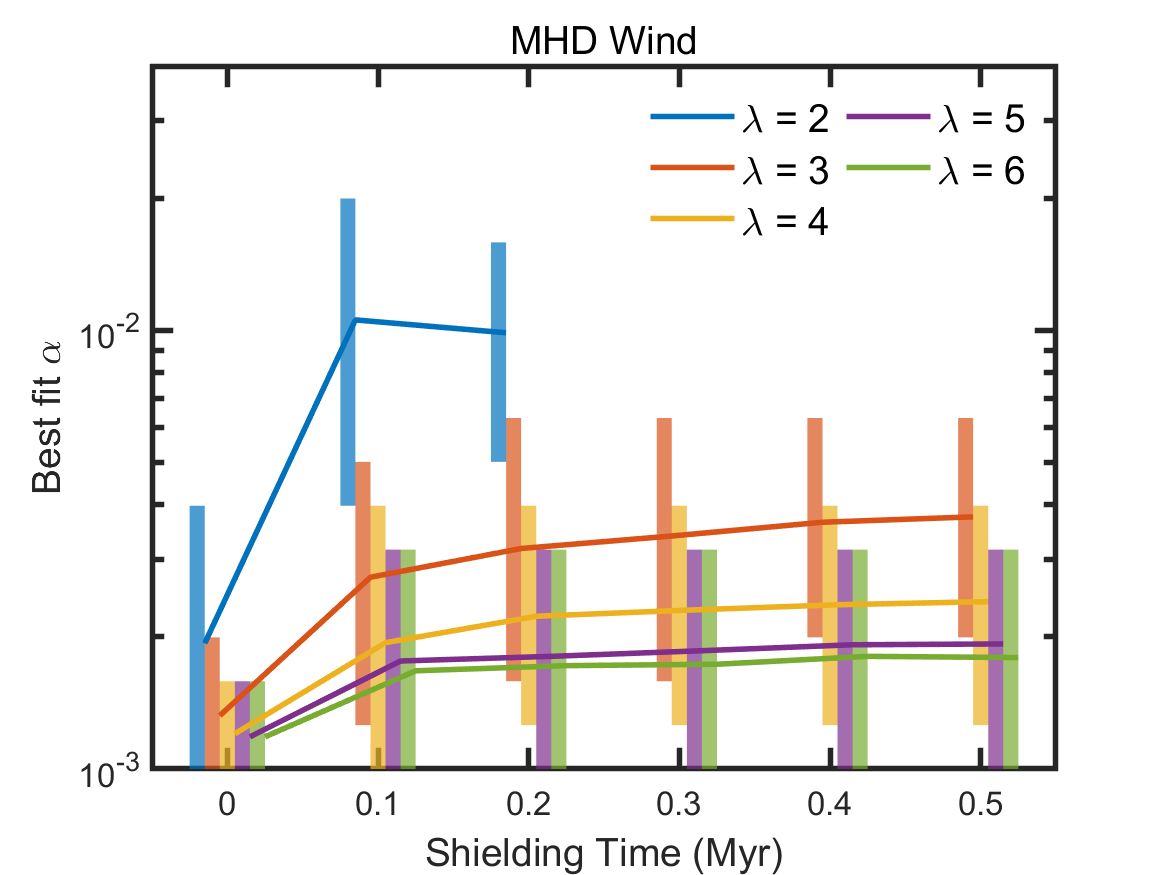}
\includegraphics[scale=0.55]{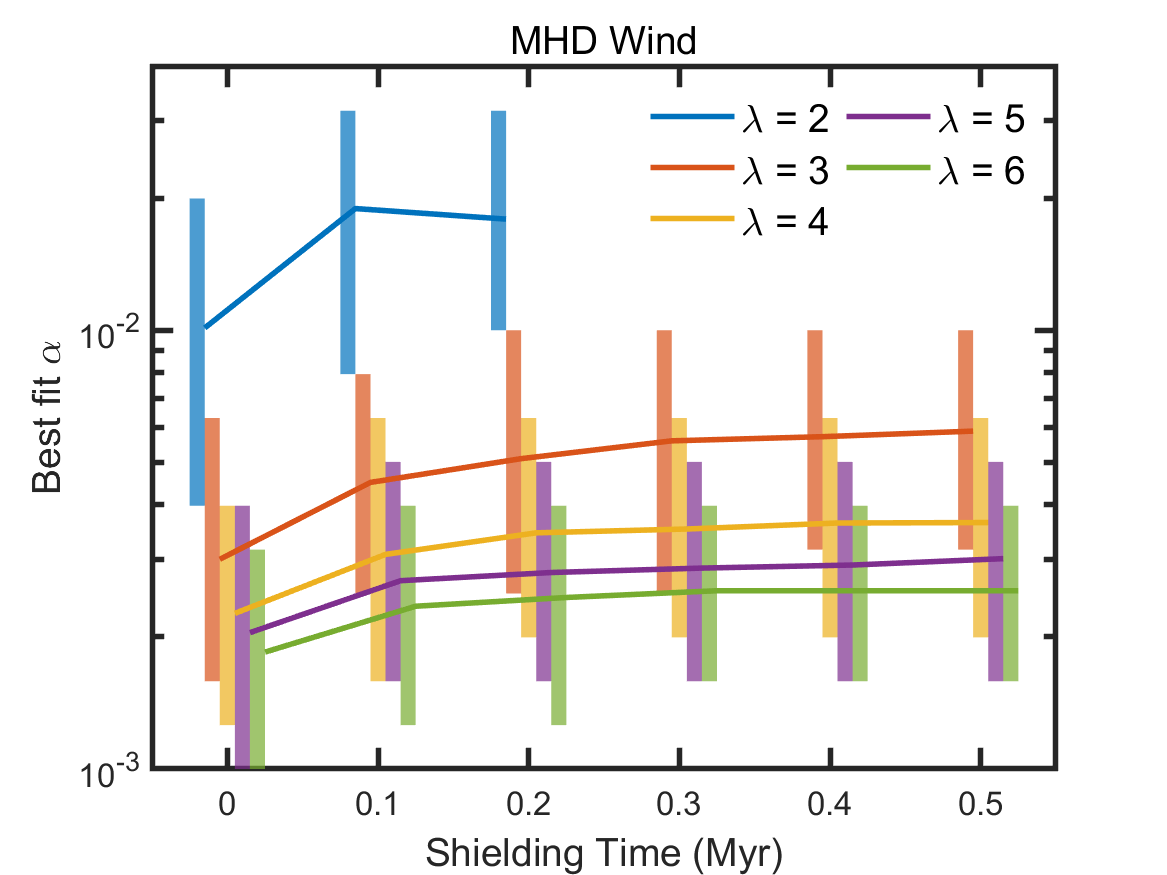}
\caption{The best fit values for $\alpha$ for MHD discs with varying values for the magnetic lever arm $\lambda$. The left-hand panel shows the best fit values for discs with initial scale radius $r_{\rm C}=20\au$, with the right-hand panel showing the values for $r_{\rm C}=50\au$.}
\label{fig:lever_arm}
\end{figure*}

Here we explore the effects of the magnetic lever arm $\lambda$ on the best fit values that arise from the MHD wind driven discs. With the magnetic lever arm $\lambda$ quantifying the efficiency of the wind at carrying away angular momentum, the lower the value of $\lambda$, the more mass that is required to be lost in the wind in order to drive accretion through the disc. In the simulations above, we used $\lambda=3$, consistent with observations that found $\lambda\simeq2$--6 \citep{Tabone17,DeValon20,Booth21,Nazari24}. To determine the effect of different magnetic lever arm values, we run an additional set of simulations, similar to those in Sect. \ref{sec:wind}, where we vary $\lambda$ to be equal to 2, 3, 4, 5 or 6. We only explore the extremes values of the initial scale radius of 20 and 50 $\au$. Figure \ref{fig:lever_arm} shows the best fit $\alpha$ values as a function of the shielding time for these simulations, with the left-panel showing them for $r_{\rm C}=20\au$, and the right-hand panel showing for $r_{\rm C}=50\au$. The red line in each case ($\lambda=3$) corresponds to the same results as in the right-hand panel of Fig. \ref{fig:best_alpha}. It is clear though from Fig. \ref{fig:lever_arm} that as the lever arm increases, the best fit $\alpha$ values decrease but quickly stabilise at similar levels of 1--4$\times 10^{-3}$ for discs with $r_{\rm C}=20\au$, and 2--6$\times 10^{-3}$ for discs with $r_{\rm C}=50\au$. The exception to this is for when $\lambda=2$, where a larger amount of material is extracted by the wind in regions of the disc close to the star. This results in larger values of $\alpha_{\rm DW}$ being required in order to match the observed accretion rate since the gas surface density is significantly reduced. Ultimately, the variations in the magnetic lever arm indicate the those values must be larger than 3 in order to match the observables for a wide range of ages, or the disc must be very young, i.e. $\le0.2$Myr, and contain a large value for $\alpha_{\rm DW}$ if the magnetic lever arm is small, e.g. 2.

\label{lastpage}
\end{document}